\documentclass[10pt, conference, compsocconf]{IEEEtran}

\usepackage{pifont}
\usepackage{cite}
\usepackage{multirow}
\usepackage[dvips]{graphicx}
\usepackage{algorithm}
\usepackage{algorithmic}
\usepackage{color}
\def\Min{{\rm minimize}}
\def\Max{{\rm maximize}}
\def\ST{{\rm subject\ to }}
\def\indif{{\tt d}}

\def\bydef{\stackrel{\triangle}{=}}
\def\beq{\begin{equation}}
\def\eeq{\end{equation}}

\def\f{{\textbf f}}

\def\bc{{\bf c}}
\def\bd{{\bf d}}

\def\bn{{\bf n}}

\def\bx{{\bf x}}

\def\bs{{\bf s}}
\def\bw{{\bf w}}
\def\bv{{\bf v}}
\def\bp{{\bf p}}
\def\bq{{\bf q}}

\def\bB{{\bf B}}
\def\bD{{\bf D}}
\def\b0{{\bf 0}}

\newcommand\bmgnu{{\mbox{\boldmath$\nu$}}}
\def\mcD{{\mathcal D}}
\def\mcG{{\mathcal G}}
\def\mcN{{\mathcal N}}

\def\mcJ{{\mathcal J}}
\def\mcR{{\mathcal R}}

\newtheorem {theo} {\bf Theorem} [section]

\newtheorem {lem}    [theo] {\bf Lemma}

\ifCLASSINFOpdf
\else
\fi

\usepackage[cmex10]{amsmath}
\graphicspath{{figure/}}

\begin{document}


\title{One More Weight is Enough: Toward the Optimal Traffic Engineering with OSPF}

\author{
    \IEEEauthorblockN{Ke Xu$^\dag$, Hongying Liu$^\ddag$, Jiangchuan Liu$^\S$, Meng Shen$^\dag$}
    \IEEEauthorblockA{Tsinghua University$^\dag$ $\qquad$  Beihang University$^\ddag$ $\qquad$  Simon Fraser University$^\S$\\
    \{xuke,shenmeng\}@csnet1.cs.tsinghua.edu.cn  \qquad  liuhongying@buaa.edu.cn  \qquad  jcliu@cs.sfu.ca
        }
}

\maketitle

\begin{abstract}
Traffic Engineering (TE) leverages information of network traffic to generate a routing scheme optimizing the traffic distribution so as to advance network performance. However, optimize the link weights for OSPF to the offered traffic is an known NP-hard problem \cite{Fortz04}. In this paper, motivated by the fairness concept of congestion control \cite{Mo}, we firstly propose a generic objective function, where various interests of providers can be extracted with different parameter settings. And then, we model the optimal TE as the utility maximization of multi-commodity flows with the generic objective function and theoretically show that any given set of optimal routes corresponding to a particular objective function can be converted to shortest paths with respect to a set of positive link weights. This can be directly configured on OSPF-based protocols. On these bases, we employ the Network Entropy Maximization(NEM) framework \cite{PEFT} and develop a new OSPF-based routing protocol, SPEF, to realize a flexible way to split traffic over shortest paths in a distributed fashion. Actually, comparing to OSPF, SPEF only needs one more weight for each link and provably achieves optimal TE. Numerical experiments have been done to compare SPEF with the current version of OSPF, showing the effectiveness of SPEF in terms of link utilization and network load distribution.
\end{abstract}


 \ifCLASSOPTIONpeerreview
 \begin{center} \bfseries EDICS Category: 3-BBND \end{center}
 \fi

\IEEEpeerreviewmaketitle

\section{introduction}\label{sec:intro}
The primary role of Internet Service Providers (ISPs) is to guarantee
service via deploying infrastructures, managing network connectivity and balancing
traffic load inside their networks \cite{Awduche}. The goal of Traffic
Engineering (TE) is to ensure efficient routing to minimize network congestion,
so that users can experience low packet loss, high throughput, and low
latency. Traffic Engineering leverages information from traffic entering
and leaving the network to generate a routing scheme that optimizes
network performance. In particular, an ISP solves the TE problem by
adjusting the routing configuration to the prevailing traffic.

In this paper, we focus on traffic engineering within a single Autonomous Systems (AS), in which we assume that the egress point of each external destination is known and fixed. Traffic engineering thus depends on a set of performance objectives that guide path selection, as well as effective mechanisms for routers to select paths that satisfy these objectives \cite{Fortz02b}.

Open Shortest Path First (OSPF) is a commonly used intra-domain routing protocol \cite{rfcOSPF}, which provides the network operators a way to control network routing by configuring OSPF link weights. The quality of OSPF-based traffic engineering depends largely on the choice of weights. Link weights can have a reasonable default configuration based on link capacity, \emph{e.g.}, Cisco's InvCap \cite{Cisco} sets the weight of a link inversely proportional to its capacity, which can be explained by the $M/M/1$ queuing model. Although fairly intuitive and convenient, these setting approaches might lead to undesirable network load distribution, since they do not take the expected traffic demand into consideration. In practice, given network link capacities and expected traffic demands, the link weights can be optimized by ISPs according to a certain object function. However, computing the optimal link weights under the evenly traffic splitting scheme has been proven to be NP-complete \cite{Fortz04}.

\textbf{Challenges}. In this paper, we take an important step towards building an OSPF-based routing protocol that can achieve the optimal traffic engineering. Although this optimization problem has attracted a great research interest and been extensively studied (e.g., \cite{Srivastava}, \cite{Wang}, \cite{BiCriteria}), there are still several challenges to be further studied, including the following:

1. \textbf{Can we design a generic objective function to meet various providers' needs?} Network providers are usually interested in various indicators to improve the network performance in different ways, e.g.,some of them might prefer to lower the maximum link utilization, while others might try to minimize path lengths. Accordingly, various objective functions have been proposed to capture these demands. Unfortunately, a set of optimal link weights with one objective function does not necessarily perform well with another objective function, or may be even worse. Are there common features existed among these different objective functions? Can we design a generic objective function to meet providers' needs?

2. \textbf{Can we guarantee the universal existence of optimal link weights?} Preceding researches showed that the optimal link weights existed with a certain kind of object functions. Although this outcome is encouraging to some extent, we still prefer to ensure the universal existence of optimal link weights under various objective functions.

3. \textbf{Can we achieve the optimal TE for intra-domain IP networks based on OSPF?} As a distributed link-state routing protocol, OSPF uses the shortest path routing with destinations based hop-by-hop forwarding and Equal-Cost Multi-Path (ECMP) mechanism to evenly split the corresponding traffic over all available equal-cost paths. Many approaches are proposed attempting to achieve the ``optimal" routing based on OSPF. Wang et al. \cite{Wang} and Srivastava et al. \cite{Srivastava} proposed flexible solutions to efficiently split traffic over shortest paths, but these centralized solutions went against the distributed feature of OSPF. A new link-state protocol named PEFT, recently proposed by Xu et al. \cite{PEFT}, successfully realized a flexible traffic splitting scheme in a distributed manner, whereas failed to maintain the shortest paths in packet forwarding thus sacrificing a key benefit of OSPF. Guaranteeing the crucial features of OSPF in terms of scalability and efficiency are thus a great challenge in achieving the optimal traffic engineering goals based on OSPF.

\textbf{Our Approach and Contributions}. Inspired by the fairness criterion of congestion control \cite{Mo}, we firstly propose a generic objective function named $(\bq, \beta)$ proportional load balance to consider various load balance demands in TE. Then we model the optimal TE as the utility maximization of multi-commodity flows and propose a distributed dual decomposition method to compute the optimal link weights. Based on these, we develop a new OSPF-based protocol, Shortest paths Penalizing Exponential Flow-splitting (SPEF). It has been proved able to achieve the optimal TE.

Toward the optimal TE, in SPEF, we only need one more weight for each link. For simplicity, we hereafter refer to the optimal and the additional link weights as the first and second link weights, respectively. We use the Network Entropy Maximization (NEM) framework proposed in \cite{PEFT} to obtain the second link weights, aiming at maintaining the path diversity. In SPEF, packets forwarding is the same as OSPF: hop-by-hop along the shortest paths constructed based on destination according to the first link weights. When there are multiple shortest paths for some source destination pairs in view of the first link weights, the flow split ratio over the multiple shortest paths can be independently computed by the routers from the second link weights. In particular, we address the above challenges as follows:

1. To capture various operators' needs, we design a new generic objective function motivated by the fairness concept in congestion control \cite{Mo}, $(\bq, \beta)$ proportional load balance, from which a family of load balance objective functions could be derived to meet various providers' needs. E.g., $(\bq, \beta)$ proportional load balance could converge to min-max load balance with the increase of $\beta$, making the MLU minimized. It can also reduce to proportional load balance with $\beta = 1$, corresponding to maximizing the product of unused capacity in networks. The generic objective function provides a chance to make a trade-off by operators, since they can just simply vary parameter settings according to their special needs.

2. To ensure the existence of optimal link weights, we model the optimal TE as the utility maximization of multi-commodity flows with the generic objective function and theoretically show that any given set of optimal routes corresponding to a particular objective function can be converted to shortest paths with respect to a set of positive link weights, which can be explicitly formulated using the spare network capacity and objective function.

3. To achieve the optimal TE based on OSPF, we develop a new routing protocol, SPEF, proving that it can achieve the optimal TE for intra-domain IP networks. Although we leverage the NEM framework proposed in PEFT, the key difference between them is in that: SPEF realizes a flexible flow splitting over shortest paths in a distributed fashion, guaranteeing the crucial features of OSPF in terms of scalability and efficiency.

\textbf{Paper Organization.} The rest of the paper is organized as follows. We propose a new generic objective function of load balance in traffic engineering in Section \ref{sec:criteria} and theoretically prove the existence of the optimal link weights with the above generic function in Section \ref{sec:balance}. The new OSPF-based protocol is developed in Section \ref{sec:spef}, following which is the performance evaluation in Section \ref{sec:evaluation}. Related work is summarized in Section \ref{sec:related}, before we conclude with the achievements and extensions in Section \ref{sec:conclusion}.

\section{Network model and load balance Criteria}\label{sec:criteria}
The notion of load balance characterizes how traffic should be
distributed to the links. In this section, we first give the network
model and then propose some new definition for load balance, which
is motivated by the notion of fairness \cite{Kelly97}, \cite{Mo}.

\subsection{Network Model}
We consider a directed network $\mcG=(\mcN,\mcJ)$ with vertex set
$\mcN$, edge set $\mcJ$, and $R$ source-destination vertex pairs
$\{s_1, t_1\}, \cdots, \{s_R, t_R\}$. Each edge $(i,j)$ has a
capacity $c_{ij}$, which is a measure for the amount of traffic flow
it can take. A demand (traffic) for $(s_r, t_r)$ is $d_r$, which
denotes the average intensity of traffic entering the network at
vertex $s_r$ and exiting at vertex $t_r$. In the following, we use
notations $N, J$ to denote the cardinalities of sets $\mcN$ and
$\mcJ$ respectively and $\mcR=\{1,\cdots, R\}$ to denote the
source-destination vertex pairs index set.

The multi-commodity flow problem is a network flow problem with
multiple commodities (or goods) flowing through the network, with
different source and sink nodes. The more customary way to treat
routing in a network is to consider it as a multi-commodity flow
problem. Denote the destination node set with $\mcD=\{t\in \mcN:
\exists \ r\in \mcR \ {\rm s.t.} \ t_r=t\}$. The traffic flow to
each destination $t\in \mcD$ can be regarded as a commodity. The
flow of commodity $t$ along edge $(i,j)$ is $f^t_{ij}$. Find an
assignment of flow satisfying  the constraints:
\begin{subequations} \label{eq:MCFC}
\begin{align}
 & f_{ij}\bydef \sum_{t\in \mcD} f^t_{ij}\le c_{ij}, \ \ \forall (i,j)\in \mcJ  \label{eq:MCFCCC}\\
 &\sum_{j:(s,j)\in \mcJ}f^t_{sj} - \sum_{i:(i,s)\in \mcJ}f^t_{is} = d_s^t, \forall t\in \mcD, \forall s\in \mcN \backslash \{t\},  \label{eq:MCFCFC} \\
 & \ f_{ij}^t\ge 0,  \ \forall t\in \mcD,  (i,j)\in \mcJ, \label{eq:MCFCNN}
\end{align}
\end{subequations}
where \eqref{eq:MCFCCC} and \eqref{eq:MCFCFC} are the capacity
constraints and flow conservation constraints, respectively, and
$d_s^t\ge 0$ is the expected traffic entering the network at node
$s$ and destined to node $t$.  Set $d_s^t=d_r$ if there exists $r\in
\mcR$ such that $s_r=s$ and $t_r=t$, or set $d_s^t=0$ otherwise.

We say a traffic distribution $\f=(f_{ij},(i,j)\in \mcJ)$ is
feasible if there exists $(\f^t,t\in\mcD)$ such that $(\f, \f^t,
t\in \mcD)$ satisfies the multi-commodity flow constraints
\eqref{eq:MCFC}. If $\f$ is feasible, the total load on and the
utilization of the link $(i,j)\in \mcJ$ are $f_{ij}$ and
$\frac{f_{ij}}{c_{ij}}$ respectively, which depend on how the
network decides to route the traffic. Now, one main task is to find
a \emph{appropriate} and \emph{feasible} traffic distribution $\f$.

An objective function enables quantitative comparisons between
different routing solutions in terms of load $f_{ij}$ on the links.
Traffic engineering usually considers a link-cost function $\Phi(\f, \bc)$ that is an increasing function of $\f$. \emph{Optimal traffic
engineering} \cite{Fortz00} means that the TE cost function is minimized over multi-commodity flow constraints \eqref{eq:MCFC}.

\subsection{Load Balance Criteria}

In order to use the network resources efficiently, spare resource
(such as bandwidth) are made to ensure high probability of data
arrival to its destinations. Now we will turn to discuss the load
balance criteria based on spare link capacity for link $(i,j)$,
which is $s_{ij}=c_{ij}-f_{ij}$.

\begin{figure}[t]
\centering
\includegraphics[width=5cm,height=1.5cm]{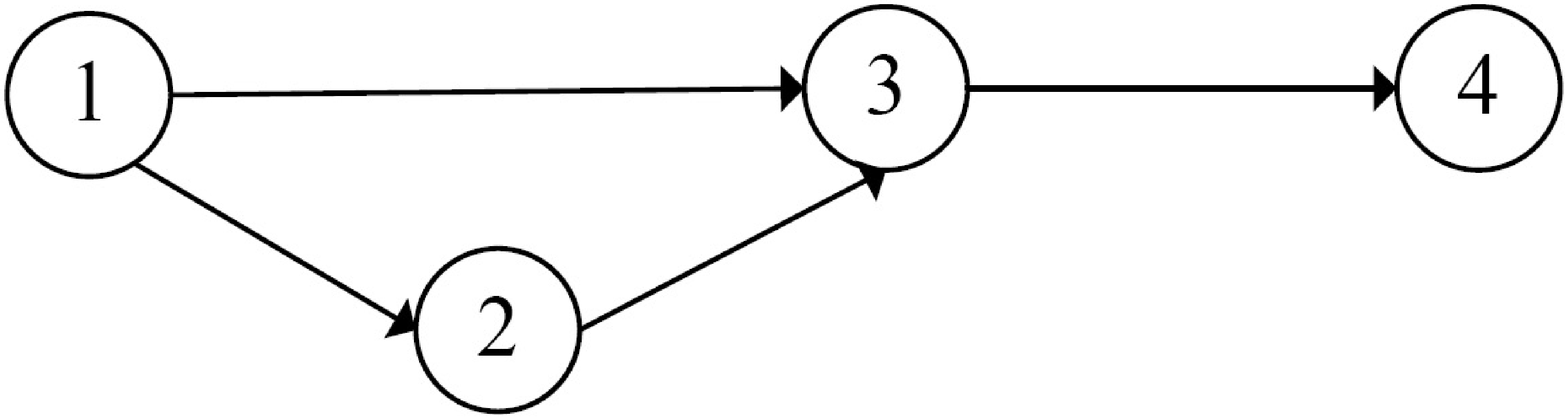}\\
\renewcommand{\figurename}{Fig.}
\caption{An example illustrating the notions of load balance for TE}
\label{fig:MLUCE}
\end{figure}

\begin{table*}[!t]
\centering \caption{Weight and link utilization for different
objective functions of TE} \label{tab:BalanceNotation}
\begin{IEEEeqnarraybox}[\IEEEeqnarraystrutmode
\IEEEeqnarraystrutsizeadd{2pt}{0pt}]{x/r/Vx/r/r/c/Vx/r/r/c/Vx/r/r/c/Vx/r/r/c/Vx/r/r/c/x}
\IEEEeqnarraydblrulerowcut \\
&&&&\IEEEeqnarraymulticol{3}{t}{$\beta=0$}&&& \IEEEeqnarraymulticol{3}{t}{$\beta=1$}&&& \IEEEeqnarraymulticol{3}{t}{B. Fortz \& M. Thorup \cite{Fortz00}}&&& \IEEEeqnarraymulticol{3}{t}{min-max}&&& \IEEEeqnarraymulticol{3}{t}{MLU} \cite{Wang}& \\

&\hfill\raisebox{-3pt}[0pt][0pt]{Link}\hfill&&
\IEEEeqnarraymulticol{25}{h}{}%
\IEEEeqnarraystrutsize{0pt}{0pt}\\
&&&&\hfill {\mbox{weights}}\hfill&&\hfill {\mbox{
utilizations}}\hfill&&&\hfill {\mbox{weights}}\hfill&&\hfill
{\mbox{utilizations}}\hfill&&&\hfill {\mbox{weights}}\hfill&&\hfill
{\mbox{utilizations}}\hfill&&&\hfill {\mbox{weights}}\hfill&&\hfill
{\mbox{utilizations}}\hfill&&&\hfill {\mbox{weights}}\hfill&&\hfill
{\mbox{utilizations}}\hfill&
\IEEEeqnarraystrutsizeadd{0pt}{2pt}\\
\IEEEeqnarraydblrulerowcut\\
&(1, 3) &&& 2&& 1.00 &&&  3   && 0.67 &&&  4.6  && 0.67 &&&  && 0.50 &&& 0&& a\textsuperscript{\ddag} &\\
&(3, 4) &&& 1&& 0.90 &&& 10   && 0.90 &&& 40.0  && 0.90 &&&  && 0.90 &&& 1&& 0.90 &\\
&(1, 2) &&& 1&& 0.00 &&&  1.5 && 0.33 &&&  2.3  && 0.33 &&&  && 0.50 &&& 0&& 1-a &\\
&(2, 3) &&& 0&& 0.00 &&&  1.5 && 0.33 &&&  2.3  && 0.33 &&&  && 0.50 &&& 0&& 1-a &\\
\IEEEeqnarraydblrulerowcut\\
&\IEEEeqnarraymulticol{7}{s}{\scriptsize \textsuperscript{\ddag}$a$
is a constant in interval $[0.1, 0.9]$}
\end{IEEEeqnarraybox}
\end{table*}

It is well known that minimizing MLU is over sensitive to
individual bottleneck links that may be difficult to avoid
\cite{Fortz02b}. In addition, the maximum link utilization function
does not penalize solutions that force traffic to traverse very long
paths. We first use an example to illustrate that MLU is not a
well-defined objective function. Consider the topology in Fig.
\ref{fig:MLUCE}, there are four edges with capacities all being 1s.
The nonzero demands are 1 for source-destination pair $(1,3)$ and 0.9 for source-destination pair $(3,4)$, respectively. There are two
paths for source-destination pairs $(1,3)$, namely 1-3 and 1-2-3.
There is a single path for source-destination pair $(3,4)$, i.e.,
3-4. The link utilizations are shown in the last column of TABLE
\ref{tab:BalanceNotation}. There is infinite optimal traffic
distribution for minimizing MLU. How to evaluate these optimal
traffic distribution? A formal definition is min-max load balance.

A traffic distribution $\f^*$ is said to be {\it min-max load
balanced} if it is feasible and for any other feasible traffic
distribution $\f$, the following condition holds: if
$s_{ij}>s^*_{ij}$ for some $(i,j)\in \mcJ$, then there exists
$(u,v)\in \mcJ$ such that $\frac{s^*_{uv}}{c_{uv}}\le
\frac{s^*_{ij}}{c_{ij}}$ and $s_{uv}<s^*_{uv}$.

We first show that a min-max load balancing traffic distribution $\f^*$ makes
MLU minimized. Minimizing MLU can be formulated with the spare
capacity as
\begin{equation}\label{eq:loadbalanceMM}
\begin{array}{ll}
 \Min & \max_{(i,j)\in \mcJ}\left(1-\frac{s_{ij}}{c_{ij}}\right)\\
\end{array}
\end{equation}
Assume that $\f^*$ is min-max load balanced and does not solve the
problem \eqref{eq:loadbalanceMM}. Then there exists a feasible
traffic distribution $\f$ such that
\begin{equation}\label{eq:MMB}
\max_{(i,j)\in
\mcJ}\left(1-\frac{s_{ij}}{c_{ij}}\right)<\max_{(i,j)\in
J}\left(1-\frac{s^*_{ij}}{c_{ij}}\right).\end{equation} Let
$(i,j)=\arg\max_{(i,j)\in
\mcJ}\left(1-\frac{s^*_{ij}}{c_{ij}}\right)$. By \eqref{eq:MMB}, we
have $1-\frac{s_{ij}}{c_{ij}}<1-\frac{s^*_{ij}}{c_{ij}}$. For $\f^*$
is min-max load balanced, there exists $(u,v)\in \mcJ$ such that
$1-\frac{s^*_{uv}}{c_{uv}}\ge 1-\frac{s^*_{ij}}{c_{ij}}$ and
$1-\frac{s_{uv}}{c_{uv}}>1-\frac{s^*_{uv}}{c_{uv}}$, which
contradicts with \eqref{eq:MMB}.

The min-max load balancing traffic distribution for the topology in
Fig. \ref{fig:MLUCE} is shown in TABLE \ref{tab:BalanceNotation}. It
can be seen that the min-max load balance is not overly sensitive to
individual bottleneck links that may be difficult to avoid. But
similar with minimizing MLU, the min-max load balancing traffic
distribution does not penalize solutions that force traffic to
traverse very long paths. Consider path 1-2-3 and path 1-3 for the
source-destination pair $(1,3)$. If the capacities in Fig.
\ref{fig:MLUCE} are five times bigger, then it would not be
worthwhile sending the traffic from $1$ through a detour over $2$ to
3. For it does not really matter that we reduce the second maximum
link utilization from $20\%$ to $10\%$.

A traffic distribution $\f^*$ is {\it proportional load balanced} if
it is feasible and for any other feasible traffic distribution
$\f$, the aggregate of proportional changes of spare capacity is
zero or negative:
\begin{equation*}\label{balancePB}
\sum_{(i,j)\in \mcJ}\frac{s_{ij}-s_{ij}^*}{s_{ij}^*}\le 0,
\end{equation*}
where $s_{ij}=c_{ij}-f_{ij}$ is the spare capacity of link $(i,j)\in
\mcJ$ for a feasible traffic distribution $\f$.

A traffic distribution $\f^*$ is \emph{weighting proportional load
balanced} if it is feasible, and if for any other feasible traffic
distribution $\f$
\begin{equation*}\label{eq:balanceWPB}
\sum_{(i,j)\in \mcJ}q_{ij}\frac{s_{ij}-s_{ij}^*}{s_{ij}^*}\le 0,
\end{equation*}
where $q_{ij}$ is a nonnegative constant for all $(i,j)\in \mcJ$.

The following definition is a generalization of proportional load
balance and min-max load balance. A traffic distribution $\f^*$ is
{\it $(\bq, \beta)$ proportional load balanced} if it is feasible and for any other feasible traffic distribution $\f$
\begin{equation}\label{eq:QbetaB}
\sum_{(i,j)\in
\mcJ}q_{ij}\frac{s_{ij}-s_{ij}^*}{(s_{ij}^*)^{\beta}}\le 0,
\end{equation}
where $s_{ij}=c_{ij}-f_{ij}$ and $\beta$ is a nonnegative parameter.

The given definition reduces to that of proportional balance with
$\beta=1$. As $\beta$ grows large, it converges to that of min-max
load balance.

\section{Optimal Weights Existence}\label{sec:balance}
In this section, we first resort to the utility maximization of multi-commodity flows to model the optimal TE. And then we theoretically prove that the optimal link weights always exist under the generic objective functions with different parameters.

\subsection{Utility Model of Traffic Engineering}\label{sec:TEU}
For the offered traffic, TE changes routing to minimize network congestion. Here we use the
utility maximization solution to route traffic which is equal to the
multi-commodity flow solution. The reason is two-fold: (a) it is optimal, \emph{i.e.}, it
gives the routing with maximum spare capacity utility; (b) it can
be realized by routing protocols that use MPLS tuneling, or in a distributed fashion by OSPF routing.

We associate link $(i,j)$ with an operator, and assume that if a spare
capacity $s_{ij}$ is held by operator $(i,j)$, which has utility
$V_{ij}(s_{ij})$ to the operator. We assume that the utility $V_{ij}(s_{ij})$ is
an increasing, concave and continuously differentiable function of $s_{ij}$ over the range $s_{ij}\ge 0$,
and $V'_{ij}(s_{ij})>0$ over the range $s_{ij}\ge 0$. Assume further that utilities are additive, so
that the aggregate utility of spare capacity $\bs=(s_{ij}, (i,j)\in \mcJ)$
is $\sum_{(i,j)\in \mcJ}V_{ij}(s_{ij})$.

It is the concavity of the function $V_{ij}$ that forces load balance
among links. If $V_{ij}$ is a convex increasing function instead of a
concave, then maximize the aggregate utility. Larger spare
capacity $s_{ij}$ should be increased, since the rate of
increase of $V_{ij}(s_{ij})$ is increasing in $s_{ij}$. When $V_{ij}$ is linear,
the rate of increase of $V_{ij}$ is the same for all $s_{ij}$.
When $V_{ij}$ is concave, a smaller spare capacity $s_{ij}$ is preferred, since
$V'_{ij}(x)>V'_{ij}(y) $ if $x<y$.

Now the optimal traffic engineering can be formulated as maximizing the aggregated utility under the multi-commodity flow constraints \eqref{eq:MCFC}.\\
TE$(V, \mcG, \bc, \bD)$
\begin{subequations}\label{eq:loadbalanceliu}
\begin{align}
 \Max_{\f^t\ge \bf 0} & \sum_{(i,j)\in \mcJ}V_{ij}(s_{ij})  \label{eq:loadbalanceliu1} \\
 \ST \quad & \bc-\sum_{t\in\mcD} {\f}^{t}  = \bs\ge {\bf 0}      \label{eq:loadbalanceliu2}\\
      & \bB\f^{t}=\bd^{t}, \ \forall t\in \mcD,             \label{eq:loadbalanceliu3}
\end{align}
\end{subequations}
where $\bB$, an $N\times J$ node-arc incidence matrix for network
$\mcG$, is introduced to represent the multi-commodity flow
constraints \eqref{eq:MCFC}. The $j$-th column of $\bB$
corresponding to link $(u,v)\in \mcJ$ is defined as
$$B_{ij}=\left\{\begin{array}{ll}
 1, & i=u\\
 -1,  & i=v\\
 0,  & {\rm otherwise,}
 \end{array}\right. $$
There is a unique optimum for the spare capacity vector
$\bs$, since the objective function \eqref{eq:loadbalanceliu1} is a
strictly concave function of $\bs$. But there may be many values of
the flow vector $(\f^t, t\in \mcD)$ satisfying relations
\eqref{eq:loadbalanceliu2} and \eqref{eq:loadbalanceliu3}. Say that
$\bs$ solves TE$(V, \mcG, \bc, \bD)$ if there exists $(\f^t, t\in
\mcD)$ such that $(\bs, \f^t,t\in\mcD)$ solves the optimization
problem \eqref{eq:loadbalanceliu}.

From the general theory of constrained convex optimization \cite{Bert99}. It follows
that $(\bs, \f^t, t\in \mcD)$ solves problem \eqref{eq:loadbalanceliu} if and
only if there exist Lagrangian multiplier vectors $\bw$ and $\bmgnu^t, t\in \mcD$ that satisfy
\begin{subequations}
\begin{align}
                  \bc-\sum_{t\in\mcD} {\f}^{t}  =&\bs,    \quad    \bB\f^{t}=\bd^{t}, \ \forall t\in \mcD  \label{eq:liuKKT1}\\
                         V'_{ij}(s_{ij})-w_{ij} =&0,      \quad    {\rm if}\ \ s_{ij}>0  \label{eq:liuKKT2}\\
                                              \le&0,      \quad    {\rm if}\ \ s_{ij}=0  \label{eq:liuKKT3}\\
                          \nu^t_j-\nu^t_i-w_{ij}=&0,      \quad    {\rm if}\ \ f^t_{ij}>0  \label{eq:liuKKT4}\\
                                              \le&0,      \quad    {\rm if}\ \ f^t_{ij}=0. \label{eq:liuKKT5}
\end{align}
\end{subequations}
we define $c'_{ij}=c_{ij}-s_{ij}$ as the \emph{target capacity} for
each link, which  is no greater than the actual capacity $c_{ij}$ (a ``virtual"
capacity). This is also desirable since it leads to an empty
equilibrium.

From $V'_{ij}(s_{ij})>0$, Eq. \eqref{eq:liuKKT2} and \eqref{eq:liuKKT3}, we have $w_{ij}>0$.
The Lagrangian multiplier vectors $\bw$ and $\bmgnu^t$ have several simple
interpretations. Let $p:v_{j_0}v_{j_1}v_{j_2}\cdots
v_{j_m}v_{j_{m+1}}$ be a possible path of source-destination pair
$(s,t)$, where $j_0=s$ and $j_{m+1}=t$. For example, if
$y_p =\min_{k=1, 2, \cdots, m+1}{f^t_{v_{j_{k-1}}
v_{j_{k}}}}>0$, we have $\sum_{(i,j)\in
p}w_{ij}=\nu^t_{t}-\nu^t_{s}\le \sum_{(i,j)\in \bar{p}}w_{ij}$ for any
other path $\bar{p}$ that connects the same source-destination pair $(s,
t)$ under the conditions \eqref{eq:liuKKT4} and \eqref{eq:liuKKT5}. We
may view $w_{ij}$ as the implied cost of traffic through link $(i,j)$.
Alternatively, $w_{ij}$ is the \emph{shadow price} of additional
capacity at link $(i,j)$. We can also regard $\bw$ as the weight set by
the operator and $\bmgnu^t$ as the vector of node potentials $(\nu^t_i: i\in \mcN)$ for destination $t$.

Let $\bw=(w_{ij}:(i,j)\in \mcJ)$. And let $(\bs, \f^t, t\in \mcD)$ be
a solution of \eqref{eq:loadbalanceliu}. We have shown that $\f^t$
determines the shortest path for each source-destination pair
$(s,t)$ under the link weights $\bw$, which is determined explicitly by the utility function $V_{ij}$ and the spare capacity
$s_{ij}$ through Eq. \eqref{eq:liuKKT2} and \eqref{eq:liuKKT3}.

If link $(i,j)$ is charged price of per unit spare capacity, and is to freely
vary the spare capacity $s_{ij}$, then the utility maximization problem for link $(i,j)$ becomes\\
Link$_{ij}(V_{ij};w_{ij})$
\begin{equation}\label{eq:loadbalanceliuLink}
\begin{array}{ll}
 \Max & V_{ij}(s_{ij})-w_{ij}s_{ij}\\
 \ST  & s_{ij}\ge 0.
\end{array}
\end{equation}
If the network receives a revenue $w_{ij}$ per unit spare capacity
from link $(i,j)$, and is allowed to freely vary the spare capacity
$s_{ij}$, then the revenue optimization problem for the network is as
follows.\\
\noindent Network$(\mcG, \bc, \bD; \bw)$
\begin{equation}\label{eq:loadbalanceliuNetwork}
\begin{array}{ll}
 \Max_{\f^t\ge \bf 0} & \sum_{(i,j)\in \mcJ}w_{ij}s_{ij}\\[1mm]
 \ST                  & \bc-\sum_{t\in\mcD} {\f}^{t}  = \bs\ge {\bf 0} \\[1mm]
                      & \bB\f^{t}=\bd^{t}, \ \forall t\in \mcD.\\
\end{array}
\end{equation}
$\bs$ solves Network$(\mcG, \bc, \bD; \bw)$ if there exists $(\f^t, t\in\mcD)$ such that $(\bs, \f^t, t\in \mcD)$ solves the problem \eqref{eq:loadbalanceliuNetwork}.

{\bf Remark 1:} Reducing the spare capacity $\bs$ from \eqref{eq:loadbalanceliuNetwork}, we have that Network$(\mcG, \bc, \bD; \bw)$ is
a minimum cost multi-commodity flow problem \cite{MCFP}, \emph{i.e.}
\begin{equation}\label{eq:MCMCF}
\begin{array}{ll}
 \Min_{\f^t\ge \bf 0} & \sum_{(i,j)\in \mcJ}w_{ij}\sum_{t\in \mcD}f^t_{ij}\\[1mm]
 \ST                  & \sum_{t\in \mcD}f^t_{ij}\le c_{ij}, \forall (i,j)\in \mcJ \\[1mm]
                      & \bB\f^{t}=\bd^{t}, \ \forall t\in \mcD.\\
\end{array}
\end{equation}

\subsection{Optimal Weights Existence}
\begin{theo}[weight-setting] \label{th:SysNetLink}
There exists a weight vector $\bw=(w_{ij}, (i,j)\in \mcJ)$ such that
the vector $\bs=(s_{ij}, (i,j)\in \mcJ)$, formed from the unique solution
$s_{ij}$ to Link$_{ij}(V_{ij};w_{ij})$, solves
Network$(\mcG, \bc, \bD; \bw)$. The vector $\bs$ also
solves TE$(V, \mcG, \bc, \bD)$.
\end{theo}

\begin{IEEEproof} Let $(\bs, \f^t, t\in \mcD)$ be a solution of TE$(V, \mcG, \bc, \bD), \bw=(w_{ij}, (i,j)\in \mcJ)$ and $(\bmgnu^t, t\in \mcD)$
be the Lagrangian multiplier vectors, i.e., Eq. \eqref{eq:liuKKT1}-\eqref{eq:liuKKT5} are satisfied. We have that
$s_{ij}$ is the unique solution $s_{ij}$ to
Link$_{ij}(V_{ij};w_{ij})$ for each $(i,j)\in \mcJ$. It can be check
that $(\bs, \f^t, t\in \mcD)$ is a KKT point of Network$(\mcG, \bc,
\bD; \bw)$ with Lagrangian multiplier vectors $(w_{ij},(i, j)\in
\mcJ)$ and $(\bmgnu^t, t\in D)$. $(\bs, \f^t, t\in \mcD)$ solves
Network$(\mcG, \bc, \bD; \bw)$, for which is a convex optimization
problem.

In addition, let a vector $(w_{ij}, (i,j)\in \mcJ)$ such that the
vector $\bs=(s_{ij}, (i,j)\in \mcJ)$, formed from the unique solution
$s_{ij}$ to Link$_{ij}(V_{ij};w_{ij})$ for each $(i,j)\in \mcJ$,
solves Network$(\mcG, \bc, \bD; \bw)$. Then there exists flow vector
$(\f^t, t\in \mcD)$ and Lagrangian multiplier vector
$(p_{ij},(i,j)\in \mcJ)$ and $(\bq^t, t\in \mcD)$ such that
\begin{subequations}
\begin{align}
     \bc-\sum_{t\in \mcD} {\f}^t = \bs,& \quad \bB\f^t=\bd^t,  \     \forall t\in \mcD    \label{eq:MCPLNetwork-KKTa}\\
                       w_{ij}-p_{ij}=0,& \quad  {\rm if}\ \ s_{ij}>0                      \label{eq:MCPLNetwork-KKTb}\\
                                 \le 0,& \quad  {\rm if}\ \ s_{ij}=0                      \label{eq:MCPLNetwork-KKTc}\\
                 q^t_j-q^t_i-p_{ij} =0,& \quad  {\rm if}\ \ f^t_{ij}>0                    \label{eq:MCPLNetwork-KKTd}\\
                                 \le 0,& \quad  {\rm if}\ \ f^t_{ij}=0.                   \label{eq:MCPLNetwork-KKTe}
\end{align}
\end{subequations}
Furthermore, by that $s_{ij}$ solves Link$_{ij}(V_{ij};w_{ij})$, we
have
$$V'_{ij}(s_{ij})=w_{ij},\ {\rm if}\  s_{ij}>0$$
and
$$V'_{ij}(s_{ij})\le w_{ij}, {\rm if}\  s_{ij}=0.$$
Replacing Eq. \eqref{eq:MCPLNetwork-KKTb} and \eqref{eq:MCPLNetwork-KKTc} by
this condition, we have that $(\f, \bs)$ satisfies conditions
\eqref{eq:liuKKT1}-\eqref{eq:liuKKT5} by replacing $\bw$ and
$\bmgnu^t$ by $\bp$ and $\bq^t$ respectively. This establishes that
$\bs$ solves TE$(V, \mcG, \bc, \bD)$, and hence the final part of
the theorem.
\end{IEEEproof}
\vspace{2mm}

\begin{figure}[t!]
\centering
\includegraphics[width=6cm,height=4cm]{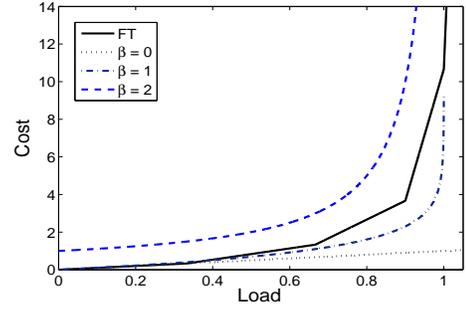}
\renewcommand{\figurename}{Fig.}
\caption{Different link cost as a function of the load for a
link capacity 1, where FT denotes the one proposed by Fortz and Thorp \cite{Fortz00} and $q_{ij}=1$ in \eqref{eq:QbetaB}} \label{fig:FTcost}
\end{figure}

We now examine the engineering implications of Theorem
\ref{th:SysNetLink}. It is true that, the Lagrangian multiplier
vector $(w_{ij}, (i,j)\in \mcJ)$  gives link weights such that all the traffic flow will be forwarded along the minimum cost multi-commodity problem solution. Meanwhile the link $(i,j)$ maximizes it's utility through retaining a proper spare capacity. Inversely, if there exists
link weights $(w_{ij}, (i,j)\in \mcJ)$ such that the vector
$\bs=(s_{ij}, (i,j)\in \mcJ)$, formed from the unique solution $s_{ij}$ to
Link$_{ij}(V_{ij};w_{ij})$ for each $(i,j)\in \mcJ$, is the same with the
solution of minimum cost multi-commodity problem \eqref{eq:MCMCF}, then $\{w_{ij}, (i,j)\in
J\}$ is a set of link weights such that all the commodity flow will
be forwarded along the shortest paths. Meanwhile, $\bs$ solves
traffic engineering problem TE$(V, \mcG, \bc, \bD)$.

\begin{figure*}
\begin{minipage}{0.46\textwidth}
  \centering
  \includegraphics[width=6cm,height=4cm]{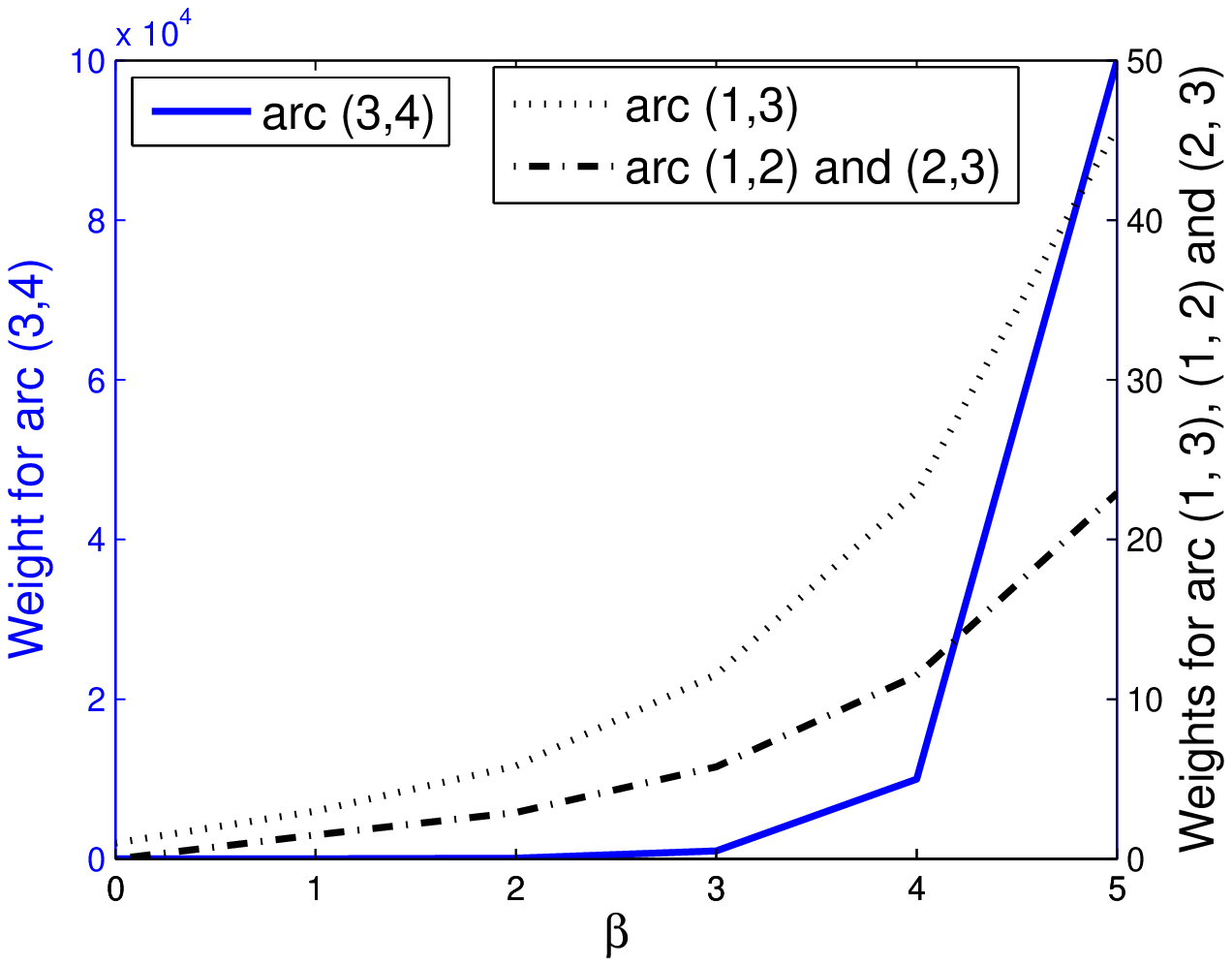}\\
  \scriptsize{(a) Weight vs load balance parameter $\beta$ with $q_{ij}=1$}
\end{minipage}\hfill
\quad
\begin{minipage}{0.50\textwidth}
\centering
\includegraphics[width=6cm,height=4cm]{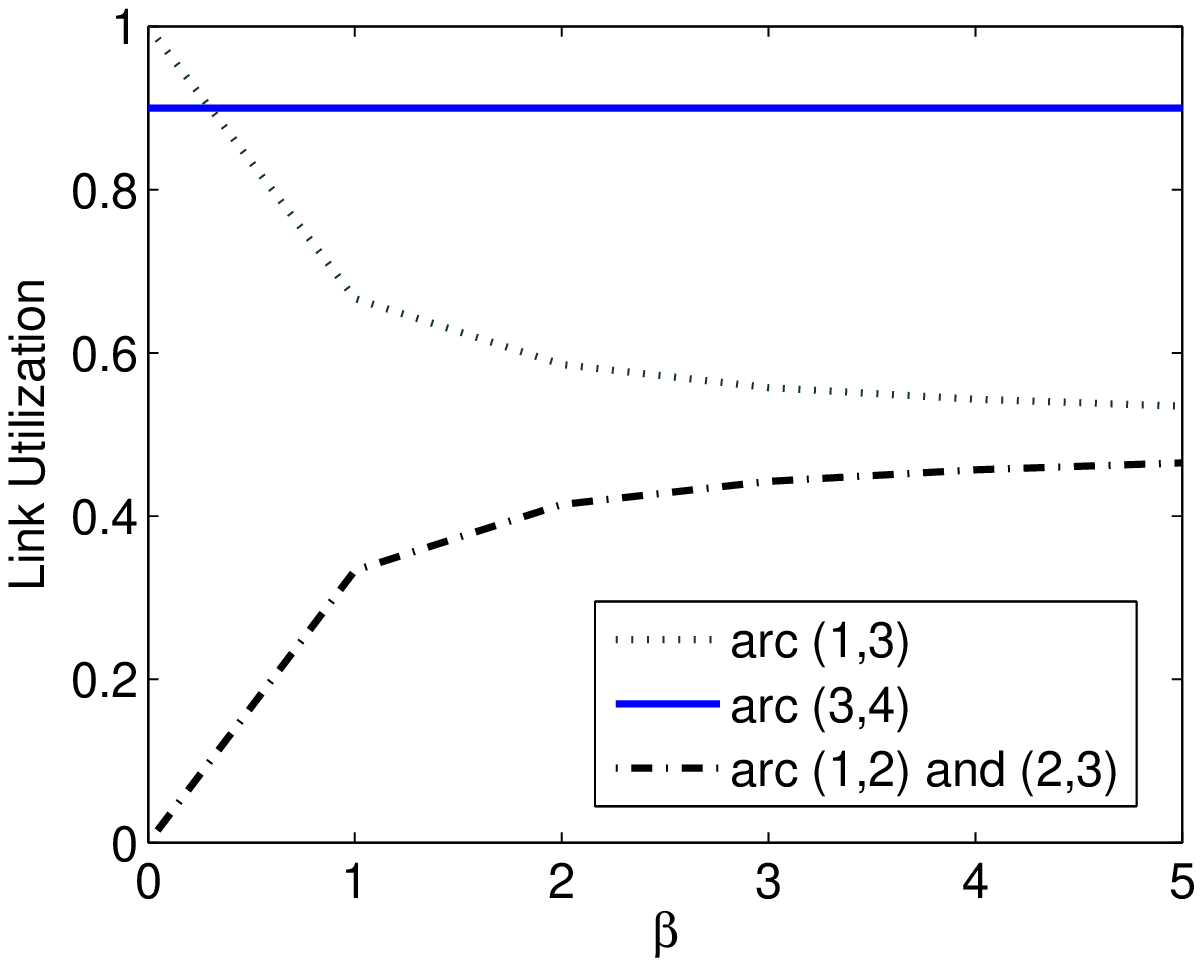}\\
\scriptsize{(b) Link utilization vs load balance parameter
$\beta$ with $q_{ij}=1$}\\
\label{fig:ToyOSPFPFET}
\end{minipage}
\renewcommand{\figurename}{Fig.}
\caption{An example illustrating the notions of load balance for TE}
\label{fig:MLUCEresults}
\end{figure*}

\begin{lem} \label{le:PBU}
Let $g(\bx)$ be continuously differentiable, i.e. $\frac{\partial
g}{\partial x_i}$ exists and continues for all $i$. It holds that
$g(\bx)$ is concave if and only if
$$g(\bx)-g(\bx^*)\le \nabla g(\bx^*)^T(\bx-\bx^*)$$
holds for any $\bx$ and $\bx^*$ (see \cite{Bert99} and \cite{Boyd}).
\end{lem}

\begin{theo} \label{th:PBU}
 A traffic distribution $\f^*$  is $(\bq, \beta)$
proportional load balanced if and only if $\bs^*=\bc-\f^*$ solves TE$(V, \mcG, \bc, \bD)$, where the objective function in \eqref{eq:loadbalanceliu1} is given by
\begin{equation}\label{eq:BalanceFunction}
V_{ij}(s_{ij}) =\left\{
\begin{array}{ll}
q_{ij}\log{s_{ij}}                    & {\rm if}\  \beta=1\\
q_{ij}(1-\beta)^{-1}s_{ij}^{1-\beta}  & {\rm if}\ \beta\neq 1.
\end{array}\right.
\end{equation}
\end{theo}

\begin{IEEEproof}  Let $g(\bs)=\sum_{(i,j)\in \mcJ} V_{ij}(s_{ij})$ and $\bs^*$ be a solution of \eqref{eq:loadbalanceliu} with objective
$g(\bs)$. Let $\f^*=\bc-\bs^*$ and $(w^*_{ij}, (i,j)\in \mcJ)$ be
a Lagrangian multiplier vector, i.e., satisfying
\eqref{eq:liuKKT1}-\eqref{eq:liuKKT5}. Let $\f$ be any feasible
traffic distribution. Then $\bs=\bc-\f\ge \bf 0$. Multiplying Eq. 
\eqref{eq:liuKKT2} and \eqref{eq:liuKKT3} by $s_{ij}-s^*_{ij}$ and
summing over $(i,j)\in\mcJ$, we find $(\nabla
g(\bs^*))^T(\bs-\bs^*)\le \sum_{(i,j)\in
\mcJ}w^*_{ij}(s_{ij}-s^*_{ij})\le 0$ since $\bs^*$ solves
Network$(\mcG, \bc, \bD; \bw^*)$ (known in the proof of 
Theorem \ref{th:SysNetLink}). Therefore,
\begin{equation}\label{eq:UtilityPB}
(\nabla g(\bs^*))^T(\bs-\bs^*)\le 0. \end{equation} By the
definition of $g(\bs)$, we have $\frac{\partial g}{\partial
s_{ij}}=\frac{q_{ij}}{s_{ij}^\beta}$. So Eq. \eqref{eq:UtilityPB}
reduces to Eq. \eqref{eq:QbetaB}. We have shown that $\f^*$ is $(\bq,
\beta)$ proportional load balance.

For the converse, assume that $\f^*$ is $(\bq, \beta)$-proportional
load balance and $\bs^*=\bc-\f^*$. So Eq. \eqref{eq:QbetaB} holds with
$\f^*$, i.e., Eq. \eqref{eq:UtilityPB}. For $g(\bs)$ is concave, we have
$g(\bs)-g(\bs^*)\le 0$ holds for any feasible traffic distribution
$\f$ such that $\bs=\bc-\f$ by Lemma \ref{le:PBU}. It follows that
$\bs^*$ solves TE$(V, \mcG, \bc, \bD)$, where the objective function
in Eq. \eqref{eq:loadbalanceliu1} is defined by 
Eq. \eqref{eq:BalanceFunction}. \end{IEEEproof} \vspace{2mm}

Based on Theorem \ref{th:SysNetLink} and Theorem \ref{th:PBU},
we can give the engineering interpretation of some specific $(\bq,\beta)$
proportional load balance.

{\bf Example 1:} Proportional load balance.

If a traffic distribution $\f$ is proportionally load balancing, then it solves the TE problem
\eqref{eq:loadbalanceliu} with $V_{ij}(s_{ij})=\log{s_{ij}}$.
From Eq. \eqref{eq:liuKKT2} and \eqref{eq:liuKKT3}, we can get
$w_{ij}=\frac{1}{c_{ij}-f_{ij}}$,
\emph{i.e.} the average packet delay on link $(i,j)$ is based on the $M/M/1$ queueing model \cite{Bert95},
where $f_{ij}=\sum_{t\in \mcD}f^t_{ij}$.
From the discussion above, we have that if path $p^*$ for $(s,t)$ bears positive traffic
$y_{p^*}>0$, then $\sum_{(i,j)\in p*}\frac{1}{c_{ij}-f_{ij}}\le
\sum_{(i,j)\in p}\frac{1}{c_{ij}-f_{ij}}$ for any other path $p$ for
$(s,t)$. The facts above show that the proportional
balance vector $\f$ \emph{not only} minimizes the average packet queueing delay
of $(s,t)$ for all $s, t\in \mcN$, \emph{but also} minimizes the average delay over all the links.

If a network is running with low utilization, then $f_{ij}\ll c_{ij}$, and therefore,
the delay $\frac{1}{c_{ij}-f_{ij}}\approx \frac{1}{c_{ij}}$. As such
InvCap recommended by Cisco can be suitable.
When the link load is non-negligible as compared to its link capacity, we should make some changes.

{\bf Example 2:} $(\bc, 2)$ proportional load balance.

If a traffic distribution $\f$ is $(\bc, 2)$ proportionally load balancing, then it solves
\eqref{eq:loadbalanceliu} with $V_{ij}(s_{ij})=\frac{-c_{ij}}{c_{ij}-f_{ij}}=-1-\frac{f_{ij}}{c_{ij}-f_{ij}}$.
In this case, we can see that \eqref{eq:loadbalanceliu} tries to minimize the total average queueing delay by the $M/M/1$
queueing model with respect to optimal link weights $w_{ij}=\frac{c_{ij}}{(c_{ij}-f_{ij})^2}$ for $(i,j)\in \mcJ$.

{\bf Example 3:} $(\bd, 0)$ proportional load balance.

Let $d_{ij}$ be the processing and propagation delay on link $(i,j)$.
If a traffic distribution $\f$ is $(\bd, 0)$ proportional load balance, then it solves the TE problem \eqref{eq:loadbalanceliu} with $V_{ij}(s_{ij})=d_{ij}(c_{ij}-f_{ij})=d_{ij}c_{ij}-d_{ij}f_{ij}$.
In this case, we can see that \eqref{eq:loadbalanceliu} tries to minimize the total processing and propagation delay, and we have that the optimal link weights $w_{ij}=d_{ij}$ for unsaturated link $(i,j)\in \mcJ$ and $w_{ij}\ge d_{ij}$ for saturated link $(i,j)$.
If $d_{ij}=1$, we have the minimum hop routing.

We use the network topology in Fig. \ref{fig:MLUCE} to illustrate these notions.
In Fig.\ref{fig:MLUCEresults}, (a) and (b) are the link weights and the
link utilization versus load balance parameter $\beta$, respectively. Detailed numerical results are shown in TABLE \ref{tab:BalanceNotation}.

{\bf Remark 2:} Consider TE$(V, \mcG, \bc, \bD)$ with $V_{ij}$
defined in \eqref{eq:BalanceFunction} with $q_{ij}=1, \forall (i,j)\in \mcJ$, we obtain: the minimum hop routing for each source-destination pair with $\beta=0$; the shortest average packet delay routing for each source-destination pair with $\beta=1$; the shortest paths for each source-destination pair which makes the traffic distribution be min-max load balance when $\beta\to \infty$.

\subsection{Utility vs. Proportional Load Balance}
In this subsection, we discuss the relation between the proportional load balance and different utility functions.
If link $(i,j)$ can choose an amount to pay per unit time, $n_{ij}$, and receive
in return a spare capacity $s_{ij}$ proportionally to $n_{ij}$, say $s_{ij}=\frac{n_{ij}}{w_{ij}}$,
where $w_{ij}$ could be regarded as a charge per unit flow for link $(i,j)$,
the utility maximization problem for link $(i,j)$ becomes\\
\noindent Link$_{ij}(V_{ij};w_{ij})$
\begin{equation}\label{eq:loadbalanceliuLink2}
\begin{array}{ll}
 \Max & V_{ij}(\frac{n_{ij}}{w_{ij}})-n_{ij}\\[1mm]
 \ST  & n_{ij}\ge 0.
\end{array}
\end{equation}
Let $\bn=(n_{ij}, (i,j)\in \mcJ), \mathcal{D}(\bn)=\{(i,j)\in \mcJ: n_{ij} >0\}$.
We define the optimization problem\\
\noindent Network$(\mcG, \bc, \bD; \bn)$
\begin{equation}\label{eq:loadbalanceliuNetwork2}
\begin{array}{ll}
 \Max_{\f^{t}\ge {\bf 0}} & \sum_{(i,j)\in \mathcal{D}(\bn)} n_{ij}\log s_{ij}\\[1mm]
 \ST                      &\ \bc-\sum_{t\in\mcD} {\f}^{t} =\bs\ge {\bf 0}  \\[1mm]
                          & \ \bB\f^{t}=\bd^{t}, \ \forall t\in \mcD.
\end{array}
\end{equation}

Note that if $n_{ij}=1$ for $(i, j)\in \mcJ$, then the solution to
Network$(\mcG, \bc, \bD; \bn)$ is the proportional load balancing traffic
allocation. If $n_{ij}, (i, j)\in \mcJ$, are all integers, then
the solution to Network$(\mcG, \bc, \bD; \bn)$ can be constructed as
follows. For each $(i,j)\in \mcJ$, replace the single link $(i,j)$ by $n_{ij}$
identical sub-links, calculate the proportional load balance
allocation over the resulting $\sum_{(i,j)\in \mcJ}n_{ij}$ traffic, and
then provide link $(i,j)$ the aggregate spare capacity allocated to
its $n_{ij}$ associated sub-links. The load \emph{per unit charge}
are then proportional load balancing.

Say that $\bs$ solves Network$(\mcG, \bc, \bD; \bn)$ if there exists
$(\f^t,t\in \mcD)$ such that $(\bs, \f^t, t\in \mcD)$ solves the optimization problem
\eqref{eq:loadbalanceliuNetwork2}.
The corresponding Lagrangian is
$$\begin{array}{ll}
 & L_{\rm Net}(\bs, \f^t, t\in \mcD; \bw,\bmgnu^t,t\in \mcD)\\[1mm]
=& \sum_{(i,j)\in \mathcal{D}(\bn)} n_{ij}\log s_{ij} +\bw^T(\bc-\sum_{t\in \mcD}\f^t-\bs)\\[1mm]
 &\quad -\sum_{t\in \mcD}{\bmgnu^t}^T(\bB\f^t-\bd^t)\\[2mm]
=& \sum_{(i,j)\in \mathcal{D}(\bn)} (n_{ij}\log s_{ij}-w_{ij}s_{ij})-\sum_{(i,j)\notin
\mcD(\bn)}w_{ij}s_{ij}\\[1mm]
 &\  +\sum_{t\in \mcD}\sum_{(i,j)\in \mcJ}(\nu^t_j-\nu^t_i-w_{ij})f^t_{ij}\\[1mm]
 &\  +\sum_{(i,j)\in \mcJ}w_{ij}c_{ij}+\sum_{t\in \mcD}{\bmgnu^t}^T\bd^t.
\end{array}$$
For the general theory of constrained convex optimization \cite{Bert99}, it follows
that $(\bs, \f^t, t\in \mcD)$ solves problem \eqref{eq:loadbalanceliuNetwork2} if and
only if there exist Lagrangian multiplier vectors $\bmgnu^t$ and
$\bw$ that satisfy:
$$
\begin{array}{rl}
            \bc-\sum_{t\in \mcD} {\f}^t = \bs,& \quad \bB\f^t=\bd^t,  \     \forall t\in \mcD \label{eq:liuNetworkKKT1}\\
                 \frac{n_{ij}}{s_{ij}}-w_{ij} &  =  0, \      {\rm if}\ \ (i,j)\in \mcD(\bn)  \label{eq:liuNetworkKKT2}\\
                                      -w_{ij} &  =  0, \      {\rm if}\ \ (i,j)\notin \mcD(\bn), s_{ij}>0 \label{eq:liuNetworkKKT3}\\
                                      -w_{ij} &\le  0, \      {\rm if}\ \ (i,j)\notin \mcD(\bn), s_{ij}=0 \label{eq:liuNetworkKKT4}\\
                       \nu^t_j-\nu^t_i-w_{ij} &  =  0, \  {\rm if}\ \ f^t_{ij}>0                  \label{eq:liuNetworkKKT5}\\
                                              & \le 0, \  {\rm if}\ \ f^t_{ij}=0.                     \label{eq:liuNetworkKKT6}
\end{array}
$$

\begin{theo} \label{th:SysNetEqui}
There exist vectors $\bn=(n_{ij},(i,j)\in \mcJ),  \bw=(w_{ij}, (i,j)\in
\mcJ)$, and $\bs=(s_{ij}, (i,j)\in \mcJ)$ such that
\begin{enumerate}
    \item [i)  ] $w_{ij}>0$ and $n_{ij}=w_{ij}s_{ij}, $ for $(i,j)\in \mcJ$;
    \item [ii) ] $n_{ij}$ solves Link$_{ij}(V_{ij};w_{ij})$ \eqref{eq:loadbalanceliuLink2}, for $(i,j)\in \mcJ$;
    \item [iii)] $\bs$ solves Network$(\mcG, \bc, \bD; \bn)$ \eqref{eq:loadbalanceliuNetwork2}.
\end{enumerate}
Given any such triple $(\bn, \bw, \bs)$, the vectors
$\bn$ and $\bs$ are uniquely determined, and $\bs$ solves TE$(V, \mcG, \bc, \bD)$.
\end{theo}

\begin{IEEEproof}
 Let $(\bs, \f^t, t\in \mcD)$ be a solution of
\eqref{eq:loadbalanceliu}. Let $\bw$ and $(\bmgnu^t, t\in \mcD)$ be the
Lagrangian multiplier vectors, i.e. the conditions
of \eqref{eq:liuKKT1}-\eqref{eq:liuKKT5} are satisfied.
 For $V'(s_{ij})>0$, we have
$w_{ij}\ge V'_{ij}(s_{ij})>0$ from \eqref{eq:liuKKT2} and
\eqref{eq:liuKKT3}. Let $n_{ij}=w_{ij}s_{ij}, (i,j)\in \mcJ$. Then i) of Theorem
\ref{th:SysNetEqui} holds.

As $w_{ij}>0$ for all $(i,j)\in \mcJ$, Link$_{ij}(V_{ij};n_{ij})$ is well
defined. In addition, it is obvious that
$n_{ij}=0$ if and only if $s_{ij}=0$. We have that $n_{ij}$ solves
Link$_{ij}(V_{ij};w_{ij})$, for $(i,j)\in \mcJ$ by
\eqref{eq:liuKKT2} and \eqref{eq:liuKKT3}. Then ii) of Theorem \ref{th:SysNetEqui} holds.

By i) of Theorem
\ref{th:SysNetEqui}, we have the fact that $\mcD(\bn^*)=\{(i,j)\in
\mcJ: n_{ij}>0\}=\{(i,j)\in \mcJ: s_{ij}>0\}$. Then the condition
\eqref{eq:liuNetworkKKT3} for Network$(\mcG, \bc, \bD; \bn)$ is
disappear. By the definition of $\bn$ and $\bw$, the conditions
\eqref{eq:liuNetworkKKT2} and \eqref{eq:liuNetworkKKT4} hold. The
other conditions follow from the conditions \eqref{eq:liuKKT1},
\eqref{eq:liuKKT4} and \eqref{eq:liuKKT5}. We have the fact that
$\bs$ solves Network$(\mcG, \bc, \bD; \bn)$. Then iii) of Theorem \ref{th:SysNetEqui} holds.

Conversely, given $\bn, \bw$ and $\bs$ satisfying
conditions i)-iii) of Theorem \ref{th:SysNetEqui}, then by
i), Link$_{ij}(V_{ij};w_{ij})$ is well defined for all $(i,j)\in \mcJ$.  In
addition, we have that $n_{ij}=0$ if and only if $s_{ij}=0$. By ii)
and $s_{ij}=\frac{n_{ij}}{w_{ij}}$, the conditions
\eqref{eq:liuKKT2} and \eqref{eq:liuKKT3} hold for given $\bs$ and
$\bw$. By iii), there exist $\f^t$ and $\bmgnu^t$ for all $t\in \mcD$ satisfying conditions
\eqref{eq:liuNetworkKKT1}, \eqref{eq:liuNetworkKKT5} and \eqref{eq:liuNetworkKKT6}. We can verify that $(\bs, \f^t, t\in\mcD, \bw, \bmgnu^t, t\in \mcD)$
satisfies \eqref{eq:liuKKT1}-\eqref{eq:liuKKT5}. We conclude $\bs$ solves
TE$(V, \mcG, \bc, \bD)$ and therefore $\bs$ is uniquely determined.
Since $w_{ij}>0$ for all $(i,j)\in \mcJ$, $\bn$ is uniquely determined as well.
\end{IEEEproof}
\vspace{2mm}

Since Theorem \ref{th:SysNetEqui} is straightforward, here we do not present the detailed proof. It shows that if each link operator is able to choose a charge per unit time prepares to pay. And if the network allocates spare capacities so that the spare capacity per unit charge is proportional load balancing, then a system optimum is achieved when the link operator's choices of charges and the network's choice of allocated spare capacities are in equilibrium.

\section{A New routing protocol: SPEF}\label{sec:spef}

We are now in a position to design a new routing protocol based on the above theoretical results. In the following, we first present the distributed algorithms to achieve the optimal link weights, also called the first link weight. And then we derive the second link weights from the conceptual framework Network Entropy Maximization \cite{PEFT}.

In the Shortest paths Penalizing Exponential Flow-splitting (SPEF), each router can construct the shortest paths for each destination based on the first link weights and independently calculate the traffic split ratio among all equal-cost shortest paths using only the second link wights, \emph{not only} achieves the optimal traffic engineering \emph{but also} remains the path diversity.

\subsection{Obtaining the First Link Weights}
We now show a distributed algorithm to obtain the first link weights, which in fact is the sub-gradient projection method \cite{Boyd} applied to the dual of TE$(V, \mcG, \bc, \bD)$. The algorithm comprises three parts: updating the weight vector, specifying the spare capacity and modifying the routing  variables, as described in \emph{Algorithm 1}.

\begin{algorithm}[H] \label{al:FirstWeight}
\caption{\emph{Dual decomposition for the first link weights}}
\small{
\begin{algorithmic}
\STATE Given tolerance $tol$ and initial weight $\bw^{(0)}$ (such as $w^{(0)}_{ij}=1/c_{ij}), k=0$;
\FOR{the given weight $\bw^{(k)}$}
    \STATE Each link $(i,j)$ solves Link$_{ij}(V_{ij};w^{(k)}_{ij})$ to find \\

    \STATE \qquad the spare capacity $s^{(k)}_{ij}$; \\

    \STATE Each destination $t\in \mcD$ solves Route$_t(\bw^{(k)};\bd^t)$:
    \STATE \begin{equation}\label{eq:loadbalanceliuRoute}
            \begin{array}{ll}
            \Min_{\f^t\ge \bf 0} & \sum_{(i,j)\in \mcJ}w^{(k)}_{ij}f^t_{ij}\\
            \ST  & \bB\f^t=\bd^t
            \end{array}
            \end{equation}
    \STATE \qquad to find the routing variable ${\f^t}^{(k)}$;\\

    \STATE Each link $(i,j)\in \mcJ$ updates the link weight
          \begin{equation}\label{eq:WeightUpdate}
            w^{(k+1)}_{ij}=\big(w^{(k)}_{ij}-\gamma_k(c_{ij}-\sum_{t\in \mcD} {f^t}^{(k)}_{ij}-s^{(k)}_{ij})\big)_+;
            \end{equation}

        \STATE $k\leftarrow k+1;$

        \STATE {\bf Until}\ $\texttt{gap}(\bw^{(k)}, \bs^{(k)}, \f^{(k)})< tol.$
\ENDFOR{}
\end{algorithmic}}
\end{algorithm}

Given the link weight $\bw$, the route problem \eqref{eq:loadbalanceliuRoute} for each destination is a minimum-cost network flow problem \cite{MCFP}. In \eqref{eq:WeightUpdate}, $\gamma_k$ is the step size and $(z)_+=\max(0,z)$. And optimality measure is defined as the dual gap, \emph{i.e.}
$$\texttt{gap}(\bw^{(k)}, \bs^{(k)}, \f^{(k)})=\sum_{(i,j)\in \mcJ}w^{(k)}_{ij}(\sum_{t\in \mcD} {f^t}^{(k)}_{ij}+s^{(k)}_{ij}-c_{ij}).$$

\begin{theo} \label{th:FirstWeight}
The link weight sequence $\{\bw^{(k)}\}$ generated by Algorithm 1 converges to the first link weights $\bw^*$ if $\sum_k \gamma_k=\infty$ and $\gamma_k\to 0$. Furthermore,  if there are no saturated links, \emph{i.e.} $s^*_{ij}>0, \forall (i,j)\in \mcJ$, the first link weights $\bw^*$ is uniquely determined and the optimal traffic distribution is $\f^*=\bc-\bs^*$, where $s^*_{ij}={V'_{ij}}^{-1}(w_{ij}^*)$.
\end{theo}
%

We have proposed a link weight configuration method that can
achieve the optimal traffic engineering. We can determine the set of
shortest paths ${\rm ON}=\{{\rm ON}^t: t\in \mcD\}$ (\emph{i.e.},
deciding which outgoing link should be chosen on the shortest path)
based on the first link weights, where ${\rm ON}^t$ is the shortest
path set for any node $s\in \mcN$ to destination $t\in \mcD$.
Specifically, ${\rm SP}^r$ denotes the shortest path set for $(s_r,
t_r)$. Let ${\rm SP}=\{{\rm SP}^r: r\in \mcR\}$. When the first link
weights generate multiple equal cost paths for a source-destination
pair or next hops for a given destination routing prefix, we need to
split the traffic among the multiple shortest paths or the next hops
to keep paths diversity while achieving the optimal traffic
engineering.

\subsection{Obtaining the Second Link Weights}
Motivated by PEFT \cite{PEFT}, we propose an Exponential-weighted
flow split in the presence of multiple equal cost paths for a
given ingress and egress pair $(s_r,t_r)$. The proposed method features that each router can
\emph{independently} compute the flow split only based on
alternative link weights, where routers can direct traffic on
the shortest paths determined by the first link weights. This method
can achieve network-wide traffic engineering  objective through OSPF
which still keeps the simplicity and scalability of link-state
routing protocols.

We maximize the relative entropy of the traffic split vector among the multipath in ${\rm SP}^r$ to maintain the path diversity.
Maximizing the relative entropy \cite{Boyd} of the traffic split vector can be formulated as follows.\\
NEM$({\rm SP}, \f, \bD)$:
\begin{subequations} \label{eq:loadbalanceOT}
\begin{align}
 \Max & -\sum_{r\in \mcR}d_r\sum_{k=1}^{n_{r}}p_k^{r}\log{p_k^{r}} \label{eq:loadbalanceOT1}\\
 \ST  &\  \sum_{r\in \mcR}\sum_{k:(i,j)\in{\rm SP}_k^{r}} d_{r}p_k^{r} \le f^*_{ij}, \forall (i,j)\in \mcJ \label{eq:loadbalanceOT3}\\
      & \sum_{k=1}^{n_{r}}p_k^{r}=1, \ \forall r\in \mcR,  \label{eq:loadbalanceOT2}
\end{align}
\end{subequations}
where $n_{r}$ denotes the number of the shortest paths from $s_r$ to $t_r$. ${\rm SP}_k^{r}$ denotes the $k$-th shortest path from $s_r$ to $t_r$.

We will connect the characterization of optimal solution to NEM
that is realizable with hop-by-hop forwarding to exponential
penalty. Let $(\bp^r, r\in \mcR)$ be a solution of
\eqref{eq:loadbalanceOT}. Then there exist Lagrangian multipliers
vector $\bv=(v_{ij}, (i,j)\in \mcJ)$ and $(\nu_r, r\in \mcR)$
satisfying that $(1+\log{p^r_k})+\sum_{(i,j)\in {\rm
SP}^r_k}v_{ij}+\frac{\nu_r}{d_r}=0, \ \forall r\in \mcR, k$ and
$\sum_{k=1}^{n_r}p_k^r=1$. Under these conditions, we have
\begin{equation}\label{eq:TrafficSplit}
p^r_k=\frac{e^{-v^r_k}}{\sum_{i=1}^{n_r}e^{-v^r_i}}, \  \forall r\in
\mcR, k,
\end{equation}
where $v^r_k=\sum_{(i,j)\in {\rm SP}^r_k}v_{ij}$. In the following, we refer to Lagrange multipliers vector $\bv$ as the second link weight.
$v^r_k$ is the length of path ${\rm SP}^r_k$ with respect to the second weight $\bv$.

\begin{theo}
The optimal traffic engineering for a given traffic can be realized with the second link weights using exponential flow split \eqref{eq:TrafficSplit}.
\end{theo}

To provide a foundation for the second link weight computation, we investigate
the Lagrange dual problem of NEM$({\rm SP}, \f, \bD)$ and a dual-gradient-based solution.
Denote the dual variables for constraints \eqref{eq:loadbalanceOT3} as $v_{ij}$ for link $(i,j)$ (or $\bv$ as a vector).
We first write the Lagrangian $L(\bp,\bv)$ associated with problem NEM$({\rm SP}, \f, \bD)$ as
$$L(\bp, \bv)=-\sum_{r\in \mcR}d_r\sum_{k=1}^{n_{r}}(p_k^{r}\log{p_k^{r}}+v_k^{r}p_k^{r})+\sum_{(i,j)\in \mcJ} v_{ij}f^*_{ij},  $$
where $v_k^{r}=\sum_{(i,j)\in {\rm SP}_k^{r}}v_{ij}$. The Lagrange dual function is
$$\begin{array}{rl}
d(\bv)= \Max & L(\bp, \bv)\\
        \ST  & \sum_{k=1}^{n_r}p_k^{r}=1, \ \forall r\in \mcR.
           \end{array}$$
The dual problem is then formulated as
\begin{equation}\label{eq:loadbalanceOTDual}
 \Min \ d(\bv) \ \ST \ \bv\ge \b0.
 \end{equation}

To solve the dual problem, we first consider the maximization of the
Lagrangian over $\bp$. Note that,  the $L(\bp, \bv)$ is separable
for a given dual variable $\bv$, \emph{i.e.}, the traffic split
subproblem for each $r\in \mcR$ is independent of the others since
they are not coupled together with link capacity constraint
\eqref{eq:loadbalanceOT3}. So we can solve a subproblem
\eqref{eq:TrafficSplitD} below for each $r\in \mcR$ separately:
\begin{equation}\label{eq:TrafficSplitD}
\begin{array}{ll}
 \Max & -d_r\sum_{k=1}^{n_r}\big(p_k^{r}\log{p_k^{r}}+v_k^{r}p_k^{r}\big)\\[1mm]

 \ST  & \sum_{k=1}^{n_{r}}p_k^{r}=1.
           \end{array}
\end{equation}
Then, the dual problem \eqref{eq:loadbalanceOTDual} can be solved by
using the gradient projection method as follows for iterations
indexed by $k$,
\begin{equation}\label{eq:GPA}
\begin{array}{rl}
 v_{ij}^{(k+1)}=&\big(v_{ij}^{(k)}-\gamma(f^*_{ij}-\sum_{r\in \mcR}d_{r}\sum_{l: (i,j)\in {\rm SP}_l^{r}} {p_l^{r}}^{(k)})\big)_+\\
         =&\big(v_{ij}^{(k)}-\gamma(f^*_{ij}-f_{ij}^{(k)})\big)_+
          \end{array}
 \end{equation}
where $\gamma>0$ is a constant step size, $({p_1^{r}}^{(k)},\cdots,
{p_{n_r}^r}^{(k)})$ are solutions of the traffic split subproblem
\eqref{eq:TrafficSplitD} for $\bv^{(k)}$, and $f_{ij}^{(k)}$ is the
total flow on link $(i,j)\in \mcJ$.

It is important to note, from \eqref{eq:GPA} in iteration
$k+1$, the procedure of link weight updating needs $f_{ij}^{(k)}$,
the aggregate bandwidth usage. We now show how to calculate it
efficiently.

\begin{table}[t!]
\renewcommand{\arraystretch}{1.3}
\caption{Forwarding table for SPEF routing.}
\label{tab:OSPF-PEFT} \centering
\begin{tabular}{c|c}
\hline                 & Lengths of multiple equal cost shortest paths through\\
      {\rm Next hop}   & link $(s,{\rm next\ \ hop})$ to $t$ in view of the second link weights \\ \hline
       $v_1$           & $(v_{11}^{(s,t)},\cdots,v_{1n_1}^{(s,t)})$       \\ \hline
       $\vdots$        & $\vdots$                                      \\ \hline
       $v_{m_s}$       & $(v_{m_s1}^{(s,t)},\cdots,v_{m_sn_{m_s}}^{(s,t)})$            \\ \hline
\end{tabular}
\end{table}

First, we need to establish the forward table for node $s$ to destination $t$ as shown in Table \ref{tab:OSPF-PEFT},
where $n_k$ denotes the number of shortest path from node $s$ through node $v_k$ to node $t$, $v_{kj}^{(s,t)}$ is the length of the $j$-th path from node $s$ through node $v_k$ to node $t$, and $m_s$ denotes the number of next hop for $s$ in ${\rm ON}^t$.
Then the traffic to destination $t$ can be splited according to the following formula:
\begin{equation}\label{eq:PEFTsplit}
\Gamma^t(s,v_k)=\frac{\sum_{j=1}^{n_k}e^{-v_{kj}^{(s,t)}}}{\sum_{i=1}^{m_s}\sum_{j=1}^{n_i}e^{-v_{ij}^{(s,t)}}},\ k=1,\cdots, m_s.
\end{equation}

Finally, the formal algorithm for the second link weights can be
described as follows, in which Algorithm $3$ is needed to get the traffic
distribution matching to the current second link weights
$\bv^{(k)}$.

\begin{algorithm}[H]

\caption{\emph{Dual decomposition for the second link weights}}
\small{
Input the optimal traffic distribution $\f^*$ and tolerance $\epsilon$;

Given the initial second link weights $\bv^{(0)}={\bf 0}, k=0$;

{\bf For} the given weights $\bv^{(k)}$, {\bf do}

\quad Get the traffic distribution matching to $\bv^{(k)}$, i.e.
$$\f^{(k)}\leftarrow {\rm TrafficDistribution}(\bv^{(k)}).$$

\quad Each link $(i,j)$ updates the second link weights
$$v^{(k+1)}_{ij}\leftarrow \big(v_{ij}^{(k)}-\gamma(f_{ij}^*-f^{(k)}_{ij})\big)_+;$$

\quad $k\leftarrow k+1;$

{\bf Until} $f^{(k)}_{ij}\le f^*_{ij}+\epsilon$ for all $(i,j)\in
\mcJ$. }
\end{algorithm}

\begin{algorithm}[H] \label{algo:TrafficDistribution}
\caption{\emph{TrafficDistribution(\bv)}} \small{ Input
${\rm ON}=\{{\rm ON}^t: t\in \mcD\}$;

Compute the path length for each path in ${\rm ON}$

\quad \ \ in view of the second link weights $\bv$;

Compute the traffic split $\Gamma^t(i,j)$ according to \eqref{eq:PEFTsplit};

{\bf For each} destination $t$ {\bf do}

\quad Do sorting on the distance of node $s$ to $t$

\quad \ \  in view of the first link weights

\quad Each source $s\neq t$ in the decreasing distance order {\bf do}

\qquad $\bar{d}_{st}=d_{st}+\sum_{(j,s)\in {\rm ON}^{t}}f_{js}^t$;

\quad {\bf For all} $j$ such that $(s,j)\in {\rm ON}^{t}$

\quad \quad $f_{sj}^t=\bar{d}_{st}\Gamma^t(s,j)$;

\quad {\bf end for}

{\bf end for}

$f_{ij}=\sum_{t\in \mcD} f^t_{ij}$ for all $(i,j)\in \mcJ$;

{\bf Return} /* set of $\f$ */}
\end{algorithm}
Here $\bar{d}_{st}$ denotes the total incoming flow destined to node $t$ at node $s$ (including traffic originating at $s$ as well as any traffic arrived from other nodes).

The following result can be proved with standard convergence analysis for gradient projection algorithms \cite{Bert99}:

\begin{theo}\label{theo:SecondWeight}
Let $\{\bv^{(i)}\}$ be the sequence generated by Algorithm 2. We have that $\{\bv^{(i)}\}$ converges to the optimal dual solutions
$\bv^*$, and the corresponding primal variables $\bp^*$ according to \eqref{eq:TrafficSplit} are the globally optimal solution of \eqref{eq:loadbalanceOT}.
\end{theo}

\vspace{2mm}


%
%
%
%
%
%
%
%
%
%
We now present a new link-state routing with hop-by-hop forwarding, which can achieve the optimal traffic engineering.

\begin{algorithm}[H]
\caption{\emph{ SPEF routing}}
\small{
 Running Algorithm 1 to obtain the first link weights $(w_{ij}, (i,j)\in \mcJ)$ and optimal traffic distribution $\f^*$.

  {\bf For} each destination node $t\in \mcD$ {\bf do}

       \quad Run Dijkstra's algorithm with the first link weights

       \quad \ \ to get all the shortest paths ${\rm ON}=\{{\rm ON}^t: t\in \mcD\}$.

       end {\bf for}.

 Running Algorithm 2 to obtain the second link weights $(v_{ij}, (i,j)\in \mcJ)$.

 {\bf For} each $t\in \mcD$ {\bf do}

 \quad {\bf For each} source node $s$:

 \quad   \quad   Establish the forward routing table shown in Table \ref{tab:OSPF-PEFT}.

 \quad {\bf end For}

 {\bf end For} }
 \end{algorithm}

\section{Performance Evaluation}\label{sec:evaluation}
How well can the new routing protocol SPEF perform? In the first part,
we will illustrate its performance with a simple example. In the second part,
we demonstrate the performance of SPEF with numerical experiments over a real backbone network and several synthetic networks. Here we compare the results of SPEF with that of OSPF, which sets link weight inversely proportional to its capacity and evenly splits the traffic over multiple equal-cost shortest paths.

\subsection{An Example}
\begin{figure}
\begin{minipage}{0.23\textwidth}
  \centering
  \includegraphics[width=4cm,height=4cm]{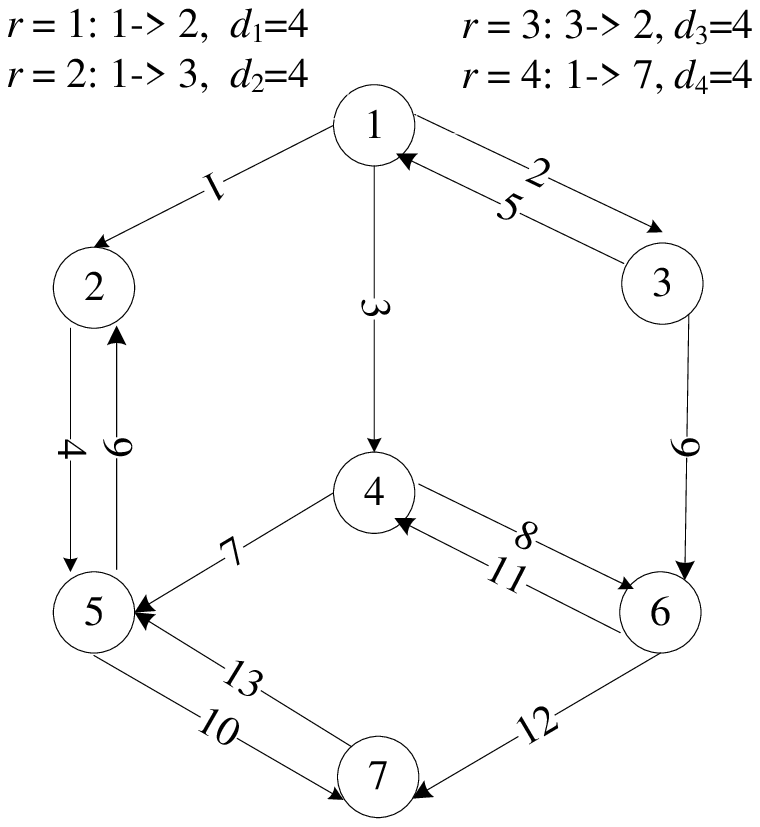}
  \renewcommand{\figurename}{Fig.}
  \caption{A simple network topology and traffic demands}
  \label{fig:Toy}
\end{minipage}\hfill
\quad
\begin{minipage}{0.25\textwidth}
\centering
\includegraphics[width=4.5cm,height=4cm]{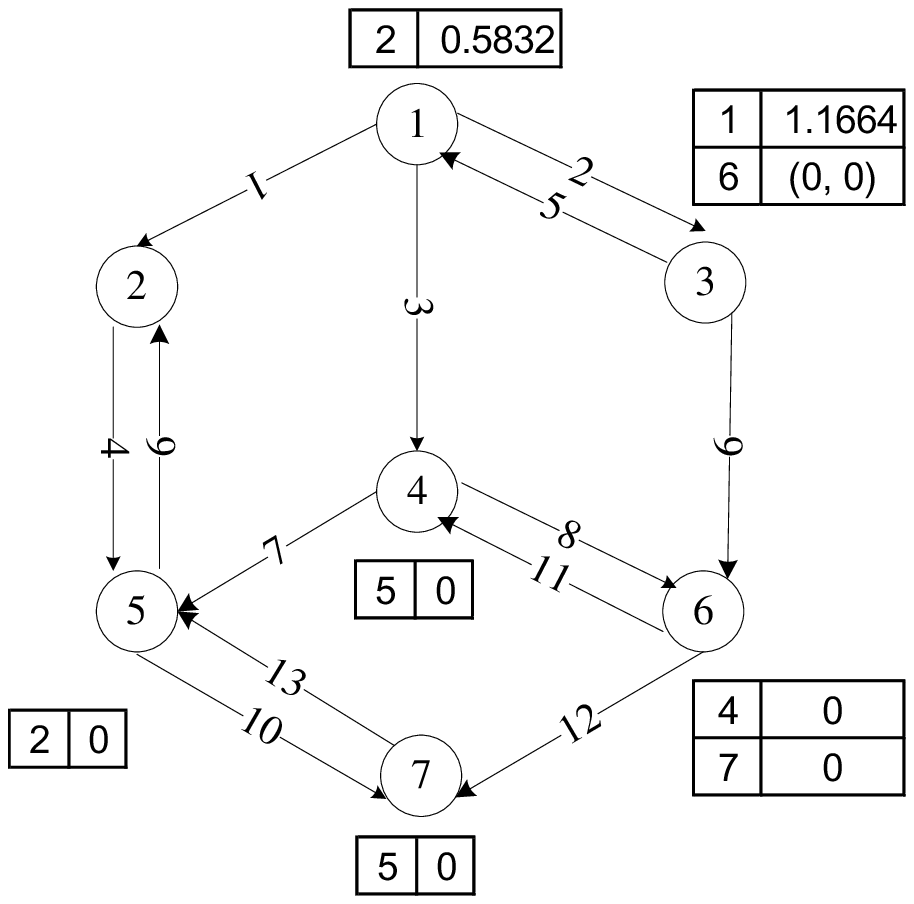}
\renewcommand{\figurename}{Fig.}
\caption{SPEF forwarding table for destination 2}
\label{fig:ToyOSPFPFET}
\end{minipage}
\end{figure}

Fig. \ref{fig:Toy} shows a simple network topology, as used in \cite{Wang}. Each link has a capacity of 5 units and each demand needs a bandwidth of 4 units. For simplicity, we omit six links unused. The numbers on the links are the link indices.

\begin{figure}[t!]
\centering
  \includegraphics[width=7cm,height=3.5cm]{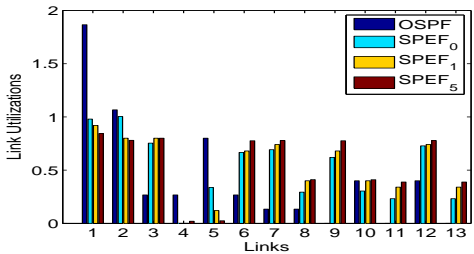}
\renewcommand{\figurename}{Fig.}
\caption{The link utilization for the topology shown in Fig.\ref{fig:Toy}}
\label{fig:ToyLinkUtilizations}
\end{figure}

\begin{figure*}[t!]
\begin{minipage}[t]{8cm}
\centering
\includegraphics[width=8cm,height=4cm]{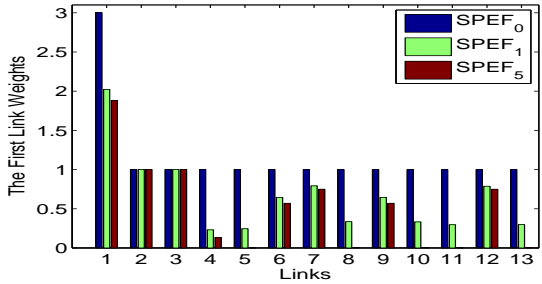}\\
\scriptsize{(a) The first link weights}\\
\end{minipage}
\quad \quad
\begin{minipage}[t]{8cm}
\centering
\includegraphics[width=8cm,height=4cm]{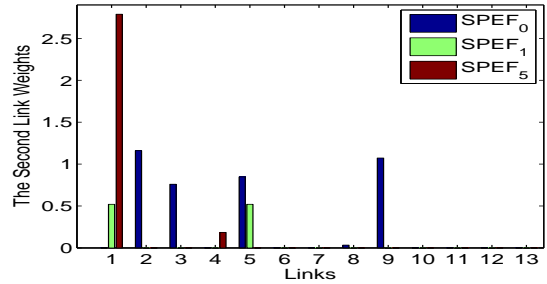}\\
\scriptsize{(b) The second link weights}
\end{minipage}
\renewcommand{\figurename}{Fig.}
\caption{The first and second link weights for the network topology shown in Fig. \ref{fig:Toy} with different $\beta$}
\label{fig:ToyFirstSecondLinkWeights}
\end{figure*}

The link utilizations for optimal TE with a different parameter $\beta$ are shown in Fig. \ref{fig:ToyLinkUtilizations}. For the results of $\beta=0$, link $1$ is a bottle link. And the first link weight is 3. The first link weight of others are all 1. Considering link 1, the link utilization is decreasing in $\beta$. From Eq. \eqref{eq:liuKKT2}, the first weight of links 2 and 3 are the same when $\beta=0,1$ or $5$, since all the spare capacities are equal to 1. For $\beta=1$, from Fig.\ref{fig:ToyFirstSecondLinkWeights} (b), it can be seen that all the second link weights are zero except for link 1 and link 5. The fact that the second weight of link 1 is increasing in $\beta$ shows we route fewer traffic through link 1 with larger $\beta$.

\subsection{Simulation Environment}

\begin{figure*}[t!]
\begin{minipage}[t]{8cm}
\centering
\includegraphics[width=6cm,height=3cm]{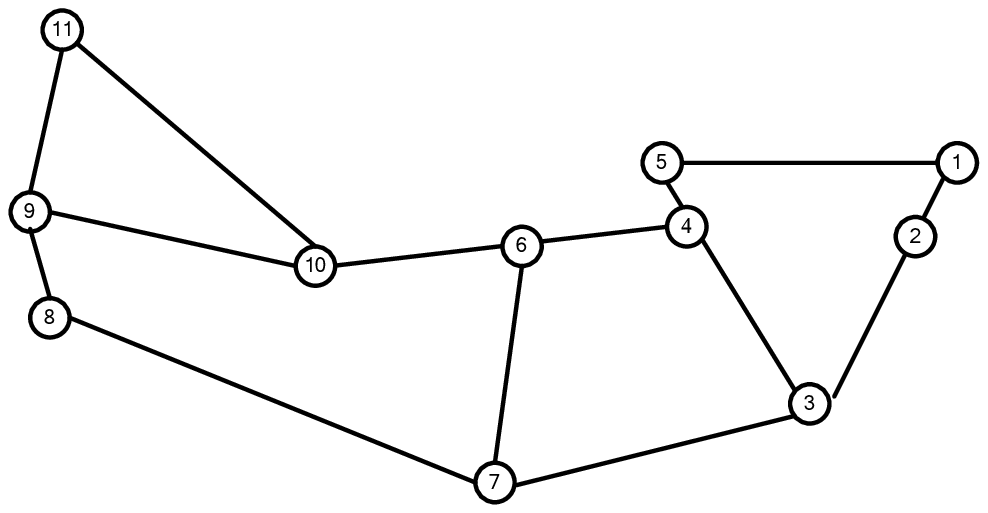}\\
\scriptsize{(a) Abilene network}
\end{minipage}
\qquad
\begin{minipage}[t]{8cm}
\centering
\includegraphics[width=7cm,height=5cm]{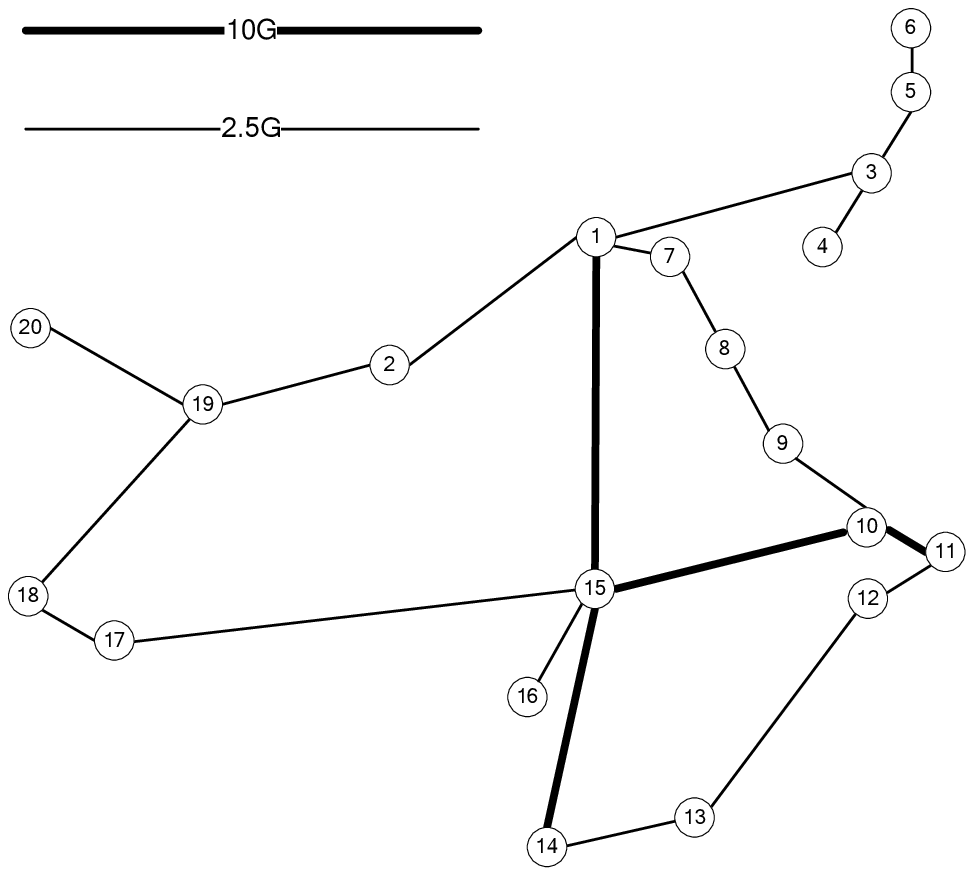}\\
\scriptsize{(b) Cernet2 network}
\end{minipage}
\renewcommand{\figurename}{Fig.}
\caption{Backbone network topologies}
\label{fig:Topology}
\end{figure*}

The properties of the networks used are summarized in TABLE \ref{tab:topology}.
The real backbone network, the Abilene network and Cernet2 network shown in Fig.\ref{fig:Topology}.
The first network has 11 nodes and 28 directional links with 10Gbps capacity, and the latter
has 20 nodes and 44 directional links with 10Gbps capacity for 4 backbone links and 2.5Gbps for others.
The traffic demands for Abilene network is generated as those in Fortz and Thorup \cite{Fortz04}.
The traffic demands for Cernet2 network are generated by a gravity model with the link aggregated load extracted from the sample Netflow data, which was captured during 2010/1/10 to 2010/1/16.
To simulate networks with different congestion levels, we create different test cases by uniformly increasing the traffic demands until the maximal link utilization almost reaches 100\% with SPEF.

We also test the algorithms proposed in this paper on the same topologies and traffic matrices as in Fortz and Thorup \cite{Fortz04}.
The 2-level hierarchical networks were generated using GT-ITM, which consist of two kinds of links: local access links with 1 unit capacity and long distance links with 5-unit capacity.
In the random topologies, the probability of having a link between two nodes is a constant parameter, and all link capacities are 1 unit.
In these test cases, for each network, traffic demands are proportionally increased to simulate different congestion levels.

For SPEF, we employ the utility function with $\beta=1$ to determine the first link weights. The utility is normalized, which means $\sum_{(i,j)\in \mcJ}\log{(1-u_{ij})}$, where $u_{ij}$ is the link $(i,j)$'s utilization. The utility is $-\infty$ if MLU is greater than 1, which is not shown in Fig. \ref{fig:UtilityFunction}.

\begin{table}[!t]
\renewcommand{\arraystretch}{1.3}
\caption{Properties for different networks}
\label{tab:topology} \centering
\begin{tabular}{||c|lrr||}
\hline
Net. ID    &  Topology  & Node \#  & Link \#   \\ \hline
Abilene    &   Backbone &    11    & 28    \\
Cernet2    &   Backbone &    20    & 44       \\  \hline
Hier50a    &   2-level  &    50    & 222      \\
Hier50b    &   2-level  &    50    & 152    \\ \hline
Rand50a    &   Random   &    50    & 242    \\
Rand50b    &   Random   &    50    & 230      \\
Rand100    &   Random   &    100   & 392       \\
\hline
\end{tabular}
\end{table}

\begin{figure*}[t!]
\begin{minipage}[t]{7cm}
\centering
\includegraphics[width=7cm,height=4cm]{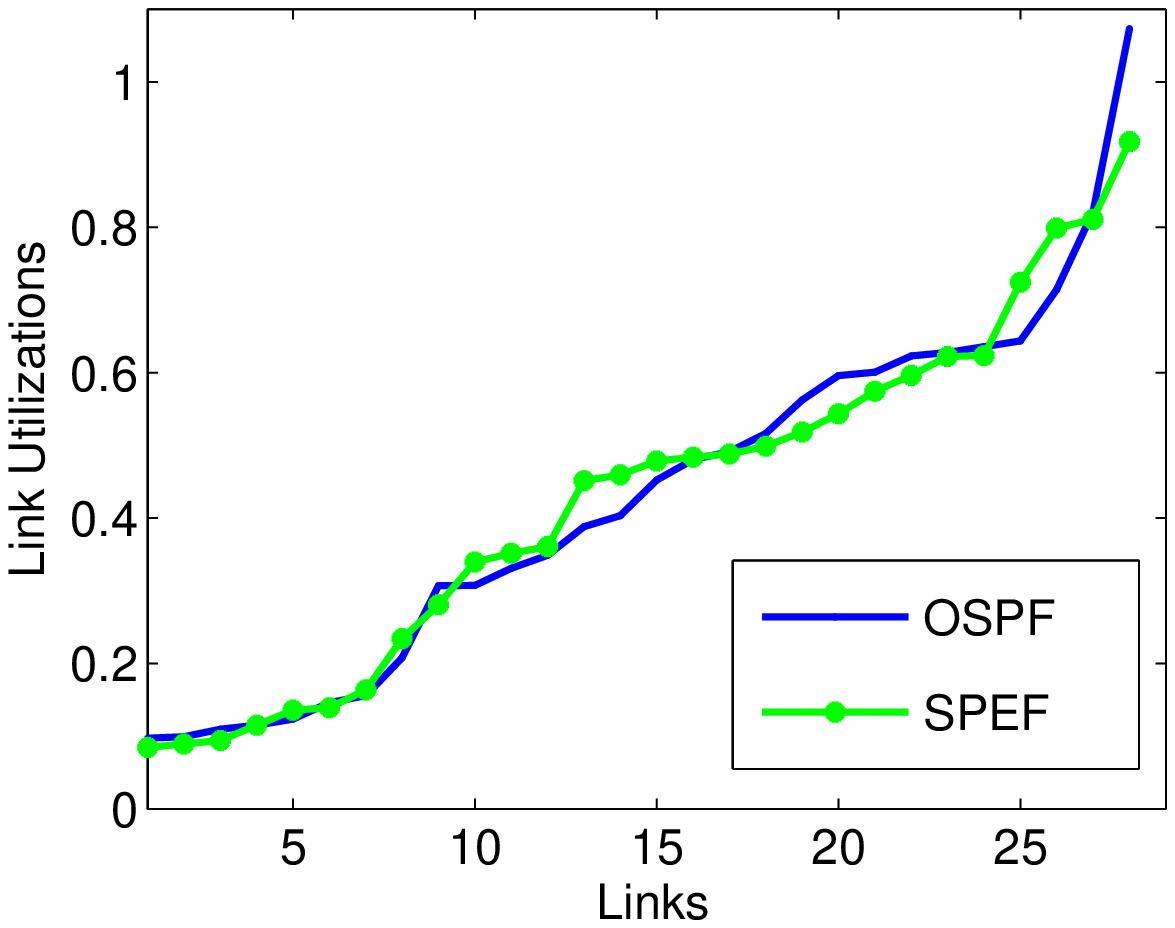}\\
\scriptsize{(a) Abilene with network load 0.17 }
\end{minipage}
\quad\quad
\begin{minipage}[t]{11cm}
\centering
\includegraphics[width=7cm,height=4cm]{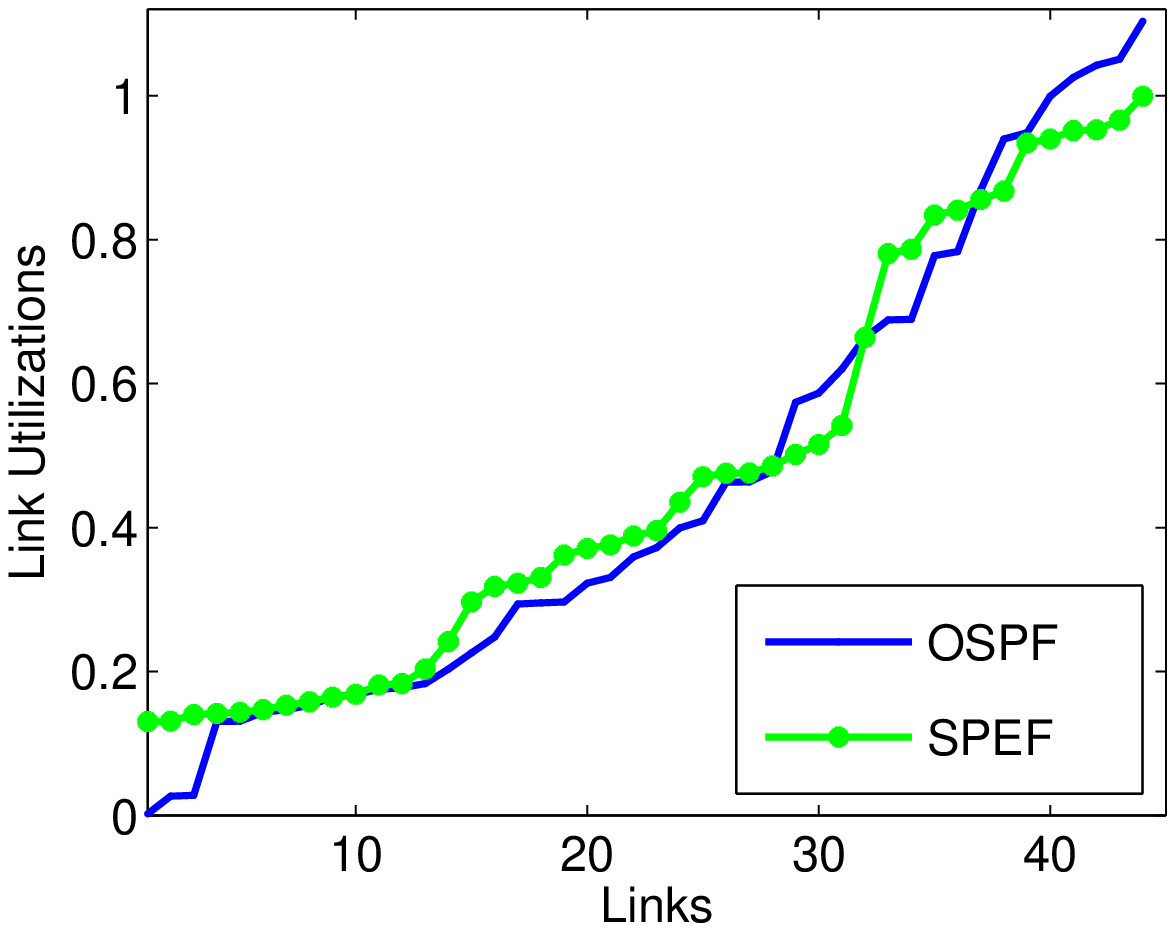}\\
\scriptsize{(b) Cernet2 with network load 0.21}
\end{minipage}
\renewcommand{\figurename}{Fig.}
\caption{Comparison of SPEF and OSPF in terms of the sorted link utilization}
\label{fig:LinkUtilization}
\end{figure*}

\subsection{Performance Comparison Against OSPF}
\begin{figure*}[t!]
\begin{minipage}[t]{4.2cm}
\centering
\includegraphics[width=4.2cm,height=3.5cm]{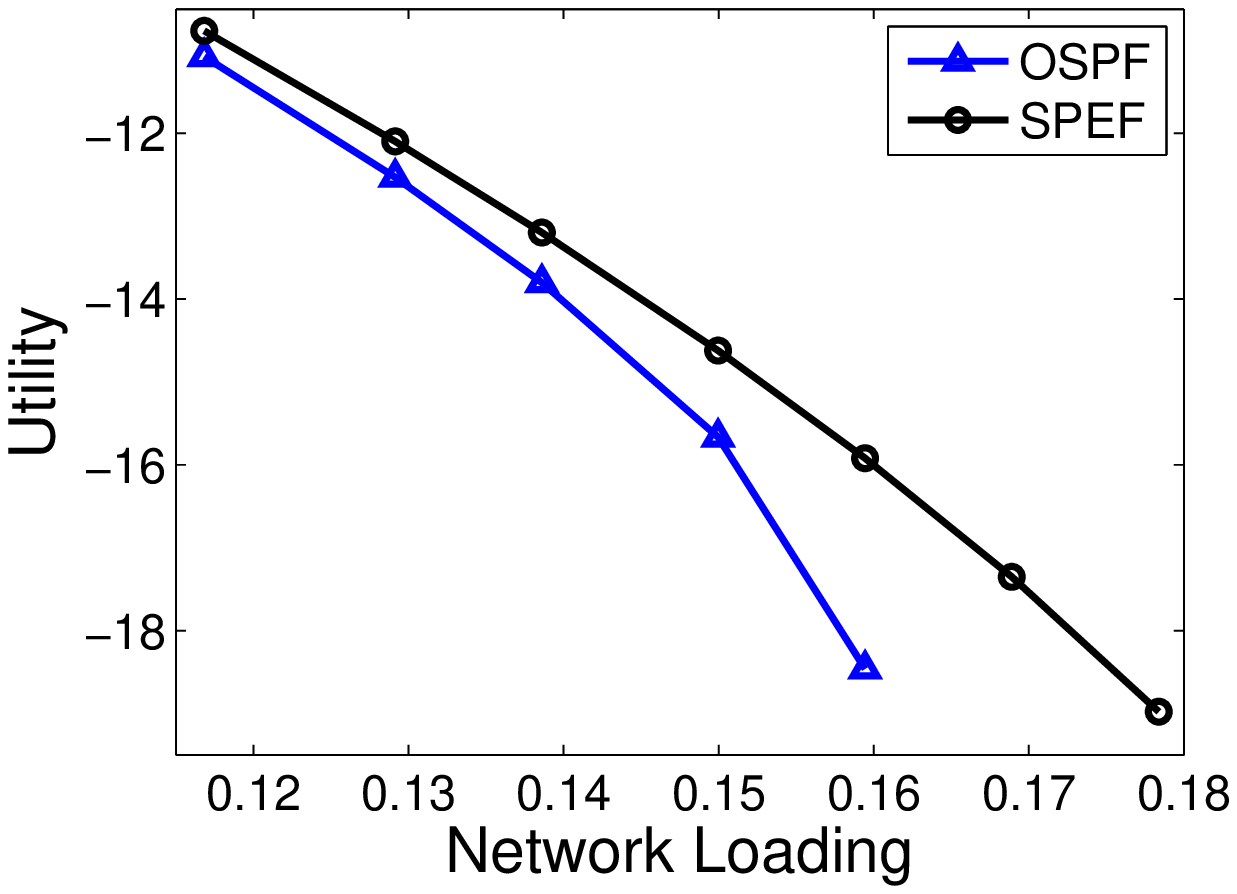}\\
\scriptsize{(a) Abilene}
\end{minipage}
\
\begin{minipage}[t]{4.2cm}
\centering
\includegraphics[width=4.2cm,height=3.5cm]{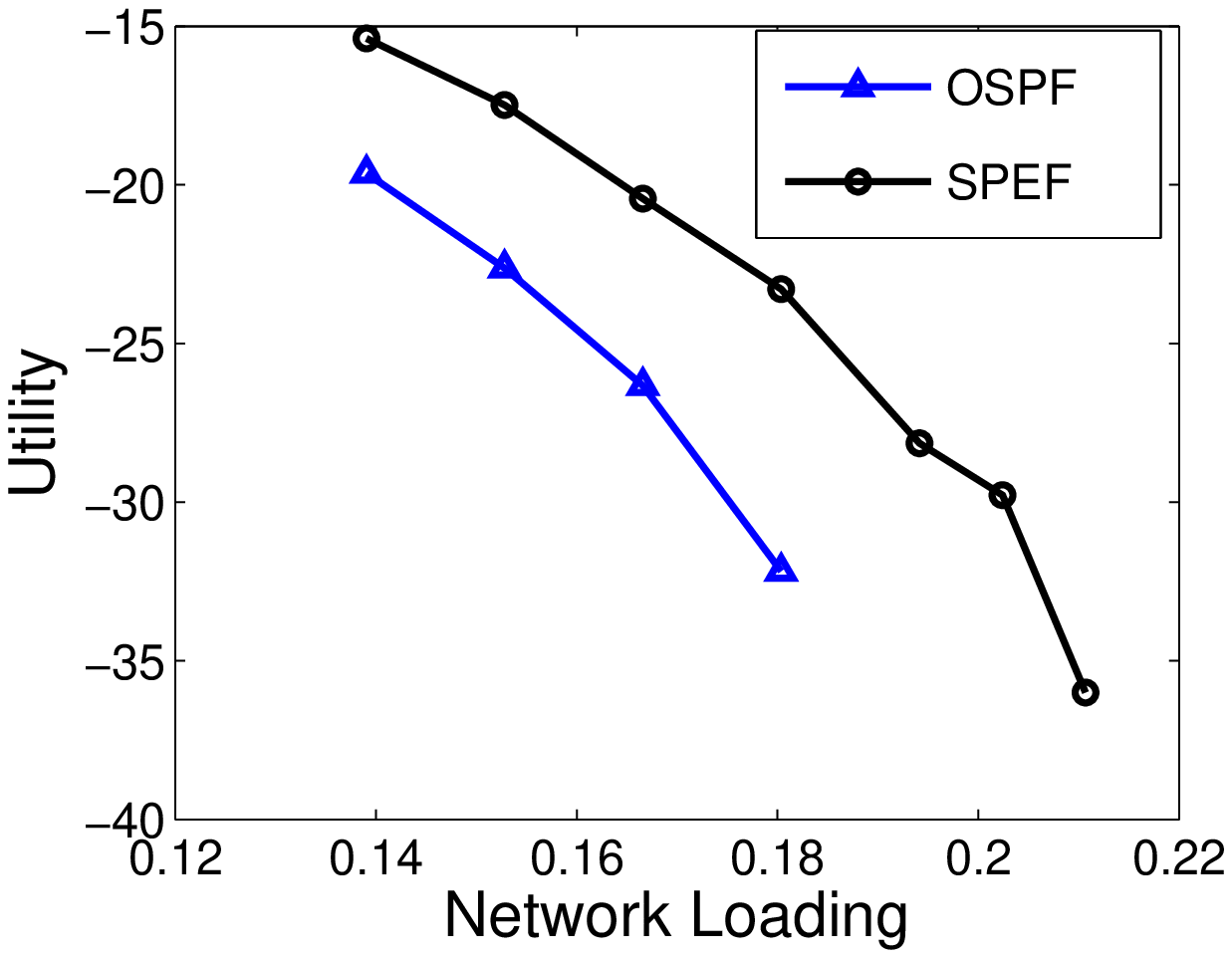}\\
\scriptsize{(b) Cernet2}
\end{minipage}
\
\begin{minipage}[t]{4.2cm}
\centering
\includegraphics[width=4.2cm,height=3.5cm]{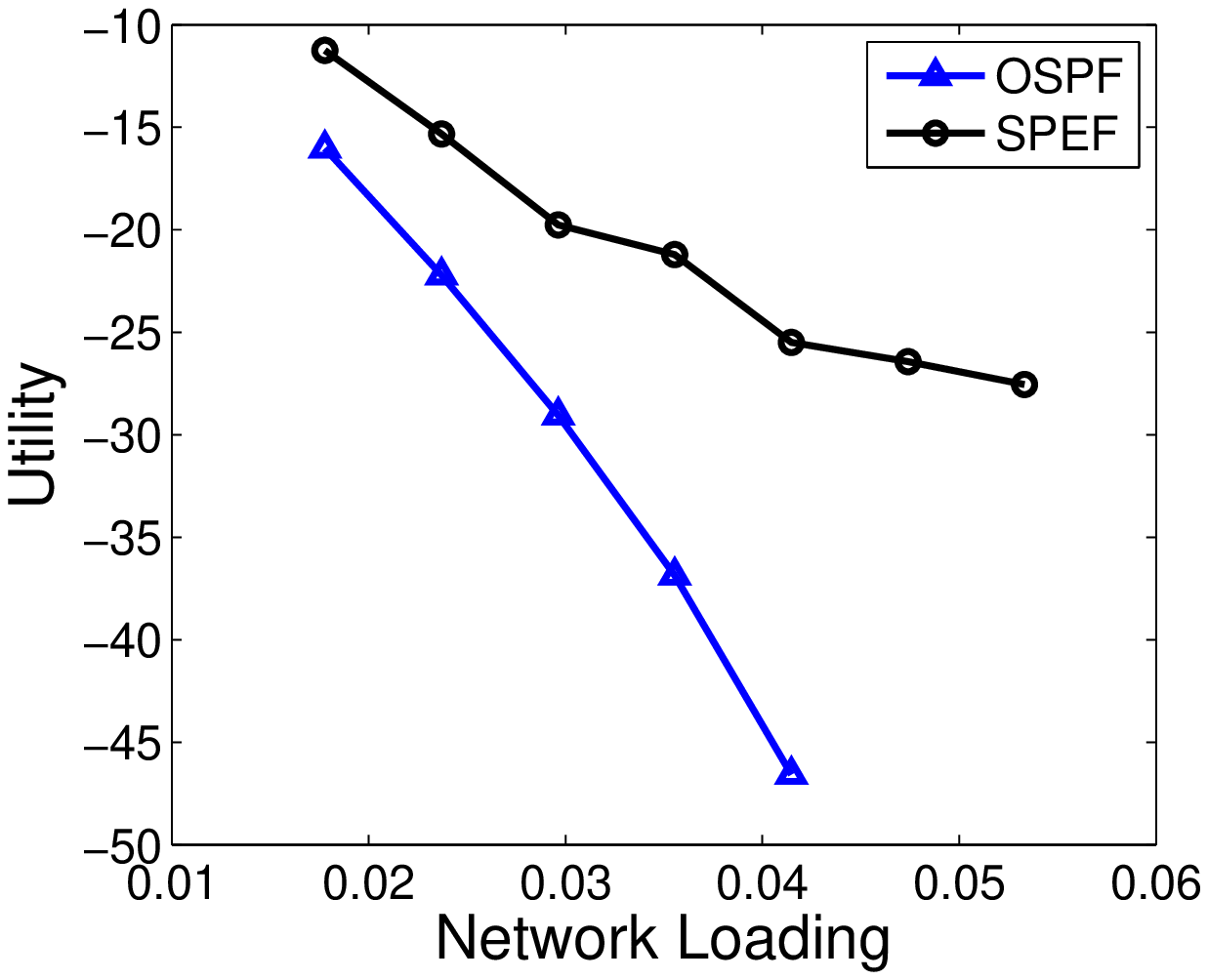}\\
\scriptsize{(c) Hier50a}
\end{minipage}
\
\begin{minipage}[t]{4.2cm}
\centering
\includegraphics[width=4.2cm,height=3.5cm]{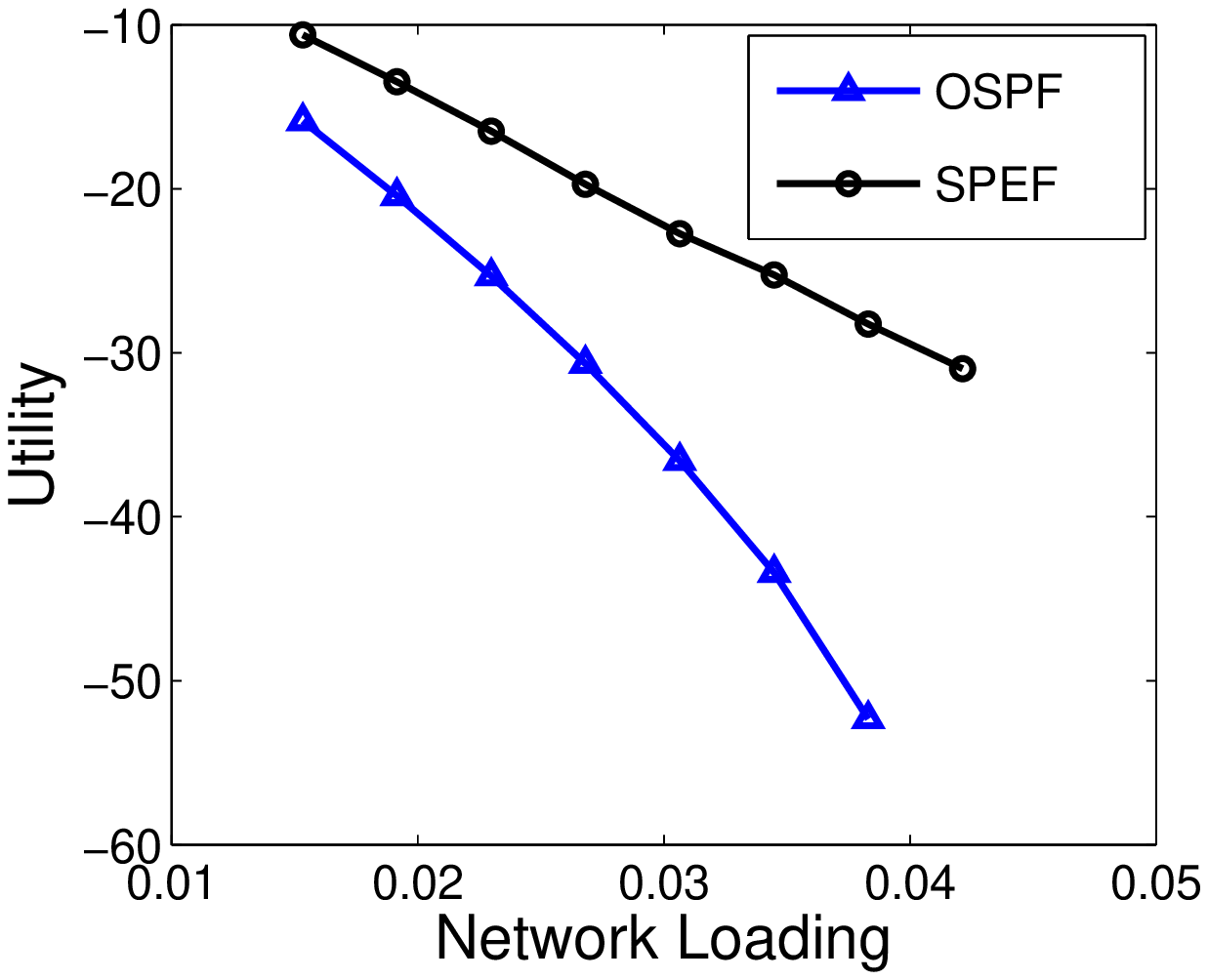}\\
\scriptsize{(d) Hier50b}
\end{minipage}

\begin{minipage}[t]{6cm}
\centering
\includegraphics[width=5cm,height=3.5cm]{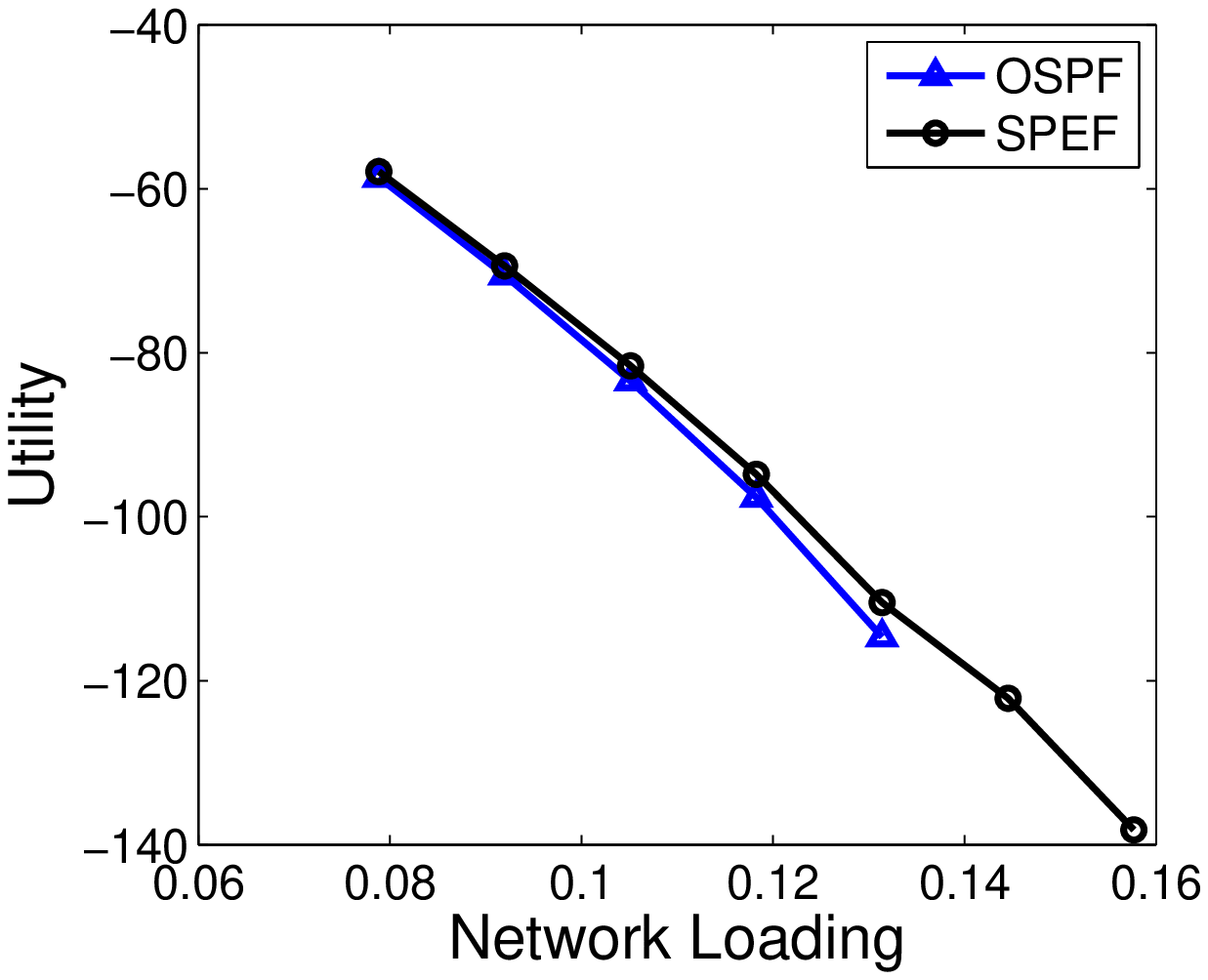}\\
\scriptsize{(e) Rand50a}
\end{minipage}
\
\begin{minipage}[t]{6cm}
\centering
\includegraphics[width=5cm,height=3.5cm]{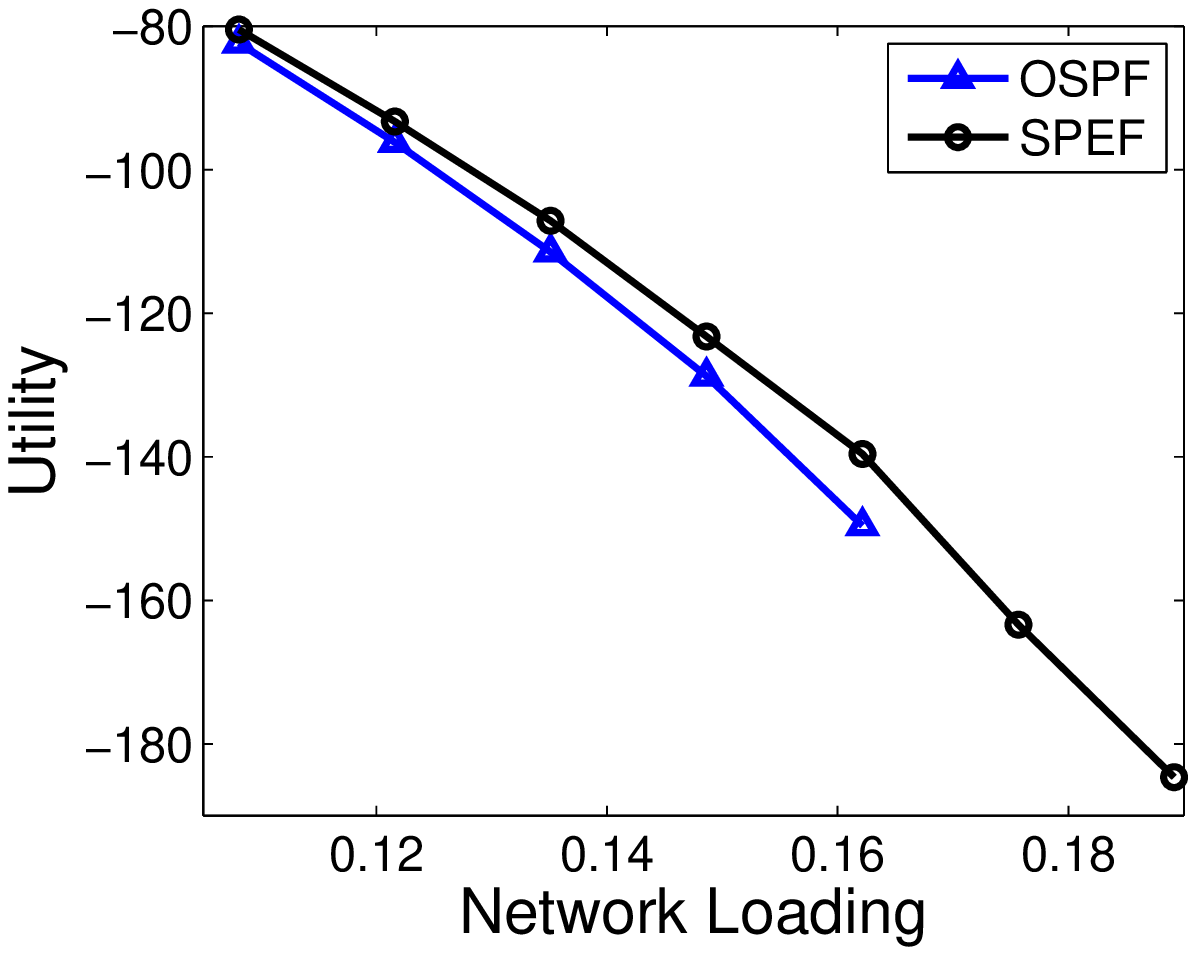}\\
\scriptsize{(f) Rand50b}
\end{minipage}
\
\begin{minipage}[t]{6cm}
\centering
\includegraphics[width=5cm,height=3.5cm]{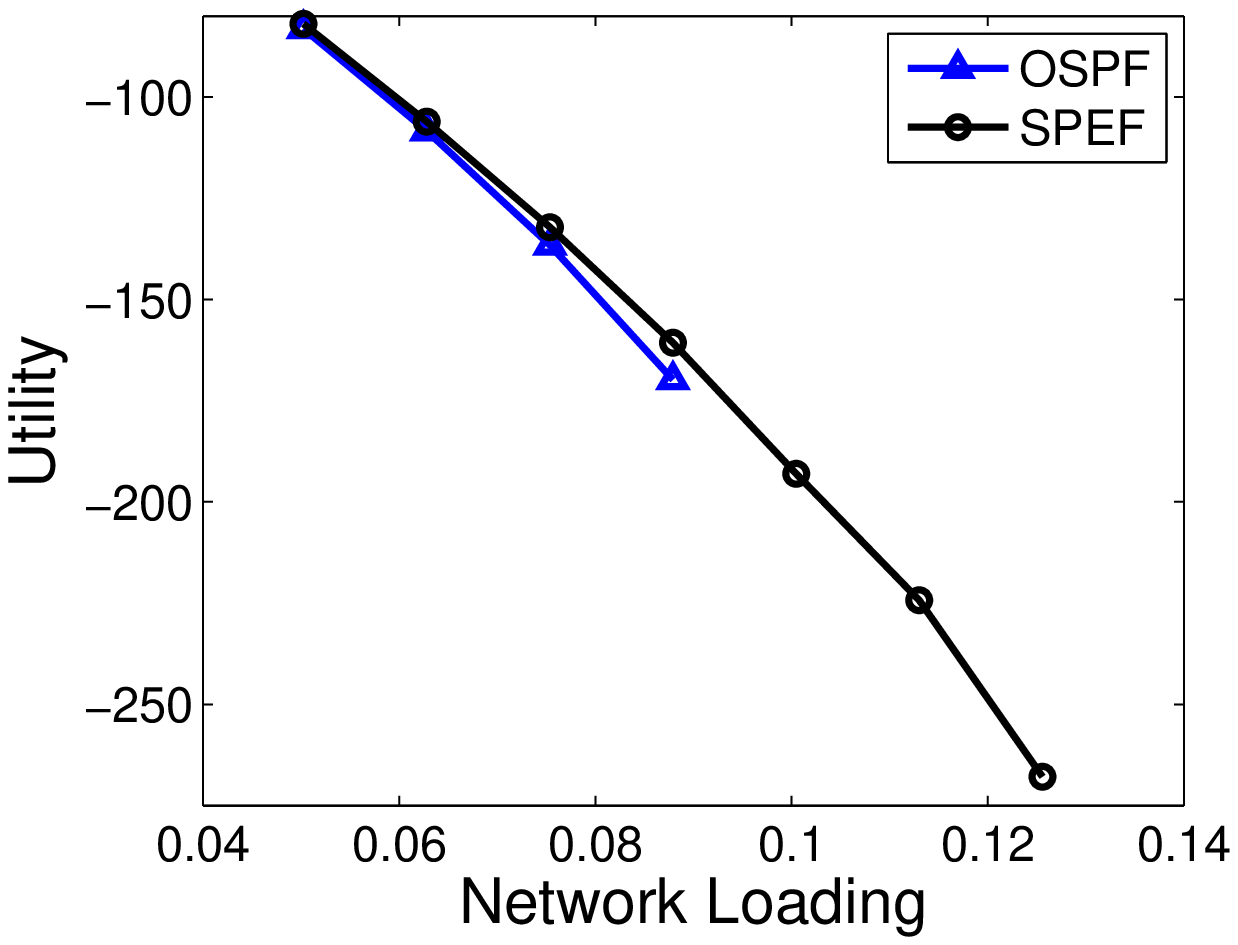}\\
\scriptsize{(g) Rand100}
\end{minipage}

\renewcommand{\figurename}{Fig.}
\caption{Comparison of SPEF and OSPF in terms of utility}
\label{fig:UtilityFunction}
\end{figure*}

The sorted link utilizations for Abilene network and
Cernet2 network are shown in Fig. \ref{fig:LinkUtilization},
where the network load is the ratio of total demand over the total
capacity. Typical results for different topologies are shown in Fig.\ref{fig:UtilityFunction}.

From Fig. \ref{fig:LinkUtilization}, it can be seen that some underutilized links in OSPF are used efficiently in SPEF. At the same time the traffic on the over-utilized links in OSPF is removed in SPEF. The results shown in Fig. \ref{fig:UtilityFunction} indicate that the utility difference between SPEF and OSPF becomes obvious with the increasing of network load.
SPEF still works when MLU of OSPF is greater than 1.

\subsection{Performance Comparison Against PEFT}
In order to make a comparison between SPEF and PEFT in network environment, we resort to SSFnet, a highly efficient simulation tool, to explore the protocol behaviors in networks with different scales. In the simulation, both the simple network in Fig. \ref{fig:Toy} and Cernet2 backbone network in Fig. \ref{fig:Topology}(b) are used, and the traffic demands are shown in TABLE \ref{tab:TrafficDemands}. Accordingly, it can be divided into two parts.

\begin{figure*}[t!]
\begin{minipage}[t]{8cm}
\centering
\includegraphics[width=6cm,height=4cm]{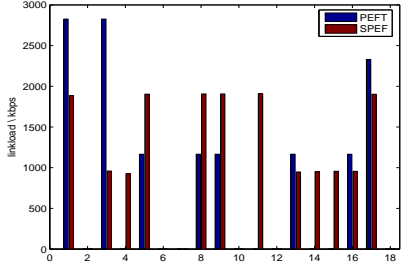}\\
\scriptsize{(a) Link load distribution over the simple network in Fig. \ref{fig:Toy}}
\end{minipage}
\qquad
\begin{minipage}[t]{8cm}
\centering
\includegraphics[width=6cm,height=4cm]{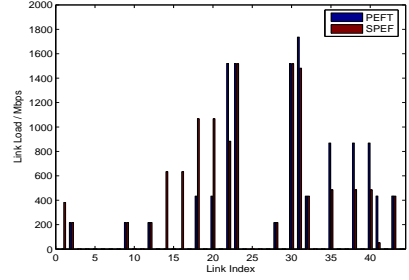}\\
\scriptsize{(b) Link load distribution over the Cernet2 network in Fig. \ref{fig:Topology}(b)}
\end{minipage}
\renewcommand{\figurename}{Fig.}
\caption{Simulation results of SPEF and PEFT using SSFnet over different networks}
\label{fig:SSFnet}
\end{figure*}

First, SPEF and PEFT separately run for 400s on the simple network with each link capacity set to be $5$Mb/s. The mean traffic load on each link is shown in Fig. \ref{fig:SSFnet}(a), of which the X axis represents link index and the Y axis is the mean traffic load in kbps. The result shows that in PEFT altogether 8 links are used for carrying traffic, and the link loads vary severely from 1000 kbps to near 3000 kbps. Comparing to PEFT, 4 more links are involved in SPEF and the traffic load is more equally distributed among these links.

Second, we run SPEF and PEFT for 400s on the Cernet2 backbone network, where the capacity of 4 links marked with bold lines is 10Gbps, four times larger than that of the rest links. The mean traffic load on each link is shown in Fig. \ref{fig:SSFnet}(b). The meaning of both X axis and Y axis are the same with those in Fig. \ref{fig:SSFnet}(a), except that the link load is measured in Mbps. Three more links is used in SPEF than that in PEFT and hence the variation of link load is lower.

In the above simulation cases, SPEF always leverages more links for packet delivery than PEFT, which can be explained by looking deeply into the forwarding tables of each protocol. Although in PEFT traffic can be intuitionally split over all possible paths between the source and destination, the penalizing exponential flow-splitting mechanism prefers the shortest path while penalizing the longer paths. In SPEF, however, multiple equal-cost shortest paths for the same source-destination pair are constructed with a higher probability based on the first link weight, and then traffic is split over these paths according to the exponential ratios computed by the second link weight. The comparison result infers that SPEF outperforms PEFT in terms of load balance.

\begin{table}[!t]
\renewcommand{\arraystretch}{1.3}
\caption{Traffic demands in comparison between SPEF and PEFT}
\label{tab:TrafficDemands} \centering
\begin{tabular}{||c|ccc||}
\hline
Net.ID.                 & Src.ID &  Dst.ID  & Demand \# \\ \hline
                        & 1 & 2 & 4Mb \\
Simple network          & 1 & 3 & 4Mb \\
in Fig. \ref{fig:Toy}   & 3 & 2 & 4Mb \\
                        & 1 & 7 & 4Mb \\
\hline
                             & 11 & 1  & 3Gb \\
                             & 11 & 2  & 2Gb \\
Cernet2 Backbone             & 11 & 20 & 2Gb \\
in Fig. \ref{fig:Topology}(b)& 13 & 6  & 1Gb \\
                             & 14 & 1  & 4Gb \\
                             & 14 & 8  & 2Gb \\
\hline
\end{tabular}
\end{table}

\subsection{Equal Cost Paths}
One of the key features of SPEF routing is the
ability to balance traffic across multiple equal-cost paths. Intuitively,
SPEF routing is more likely to use multiple paths
to balance traffic at higher loads. Hence, we focus on a different utilization scenarios for
Cernet2 network, for which we compute the
number of equal cost paths used by SPEF
routing. TABLE \ref{tab:EqualCostPath} shows the results,
where $n_i$ denotes the number of ingress-egress pairs that have $i$ equal cost paths.
It can be seen that the equal cost paths for some ingress-egress pairs are increasing with the increase of network load.
But OSPF routing has not change with the network load.

\begin{table}[!t]
\renewcommand{\arraystretch}{1.3}
\caption{Comparison of SPEF and OSPF in terms of the number of equal cost path for each ingress-egress pair}
\label{tab:EqualCostPath} \centering
\begin{tabular}{||c|c|llll||}
\hline
Routing             & Network loading &  $n_1$  & $n_2$ & $n_3$ & $n_4$   \\ \hline
OSPF & 0.13, 0.17, 0.21&  355    & 25    & 0     & 0\\  \hline
                    &   0.13          &  330    & 48    & 0     & 2 \\
SPEF &   0.17          &  325    & 53    & 0     & 2\\
                    &   0.21          &  321    & 54    & 3     & 2 \\
\hline
\end{tabular}
\end{table}

\subsection{Convergence Behavior}
\begin{figure*}[t!]
\begin{minipage}[t]{8cm}
\centering
\includegraphics[width=6cm,height=4cm]{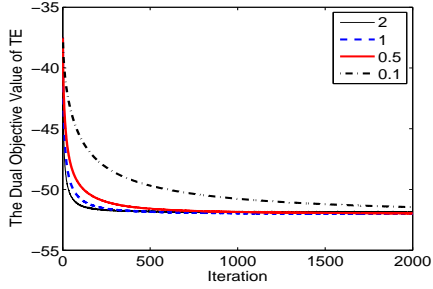}\\
\scriptsize{(a) Evolution of dual objective value of Algorithm 1}
\end{minipage}
\qquad
\begin{minipage}[t]{8cm}
\centering
\includegraphics[width=6cm,height=4cm]{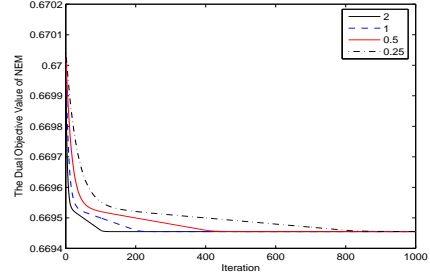}\\
\scriptsize{(b) Evolution of dual objective value of Algorithm 2}
\end{minipage}
\renewcommand{\figurename}{Fig.}
\caption{Evolution of dual objective value obtained by Algorithm 1 and Algorithm 2 with different step sizes for Cernet2 network}
\label{fig:ConvergenceBehavior}
\end{figure*}

In Algorithm 1, the initial link weights
$w_{ij}^{(0)}=\frac{1}{c_{ij}}$ for all link
$(i,j)\in \mcJ$ are a proper choose.
The step sizes in Algorithm 1 can be constant or dynamically adjusted. We find that setting the step size in Algorithm 1 to the reciprocal of the maximum link capacity $\frac{1}{\max\{c_{ij}:(i,j)\in \mcJ\}}$ performs well in practice. Fig. \ref{fig:ConvergenceBehavior} (a) shows the evolution of dual objective
value of TE obtained by Algorithm 1 with different step sizes, within the first 2000 iterations for Cernet2 network. It provides convergence behavior
typically observed. The legends show the ratio of the step size over
the default setting which is $\frac{1}{\max\{c_{ij}:(i,j)\in \mcJ\}}$. It demonstrates that Algorithm 1 developed for the SPEF routing convergence very fast with default setting.
Algorithm 1 reduces the dual objective value of TE to -48 after 100 iterations and -49 after 500 iterations. In addition, increasing step size a little will speed up the convergency, and as
expected, too large a step size (e.g., 2 for Algorithm 1 in the above example) would cause
a little oscillation. Notice that there is a wide range of step size that can make convergence very fast.

In Algorithm 2, the initial link weights $v_{ij}^{(0)}=0$ for all link $(i,j)\in \mcJ$ are a proper choose.
We find that setting the step size in Algorithm 2 to the reciprocal of the maximum optimal traffic distribution $\frac{1}{\max\{f^*_{ij}:(i,j)\in \mcJ\}}$ performs well in practice.  Fig. \ref{fig:ConvergenceBehavior} (b) shows evolution of dual objective value of NEM obtained by Algorithm 2 with different step sizes for Cernet2 network. It provides convergence behavior
typically observed. The legends show the ratio of the step size over
the default setting which is $\frac{1}{\max\{f^*_{ij}:(i,j)\in \mcJ\}}$. It demonstrates that the initial link weights for Algorithm 2 are a good approximation solution for the dual problem of NEM. And Algorithm 2 developed for the SPEF routing also convergence very fast with default setting.
Algorithm reduces the dual objective value of NEM to 0.6695 after 100 iterations and 0.66945 after 300 iterations.
In addition, increasing step size a little will speed up the convergency.
Notice that there is a wide range of step size that can make convergence very fast.

\subsection{Noninteger Link Weights}
We should point out that Algorithm 1 is not guaranteed to yield
exact integer solutions for the first link weights. In practice,
routing protocols link OSPF and IS-IS have a finite field width for
link weight information. To guarantee the weight for the link with
maximum spare capacity is 1, we get the integer weights as following
$$w'_{ij}={\rm round} \quad w_{ij}\left(\max\{s_{ij}: (i,j)\in J\}\right), \forall (i,j)\in \mcJ,$$
where ${\rm round}(x)$ means rounding $x$ to the nearest integer and $w_{ij}$ is the first link weight obtained with Algorithm 1.
This is because the modified link weights can result in routings
that are different from the optimal routing. Hence it is important
to study how errors in link weights influence performance.

There are two factors that can introduce inaccuracies in link
weights. First of course is the precision in rounding off link
weights. The second factor is the nonzero tolerance required by Dijkstra's algorithm. This implies
that the optimal link weights (and path cost) are accurate only
within a certain tolerance. For example, two path costs are treated
to be equal by Dijkstra's algorithm if the difference in costs is less
than the specified tolerance. We specify the tolerances for the Dijkstra's algorithm are 0.3 and 1 for noninteger weights and integer weights respectively.

\begin{figure*}[t!]
\begin{minipage}[t]{8cm}
\centering
\includegraphics[width=7cm,height=4.5cm]{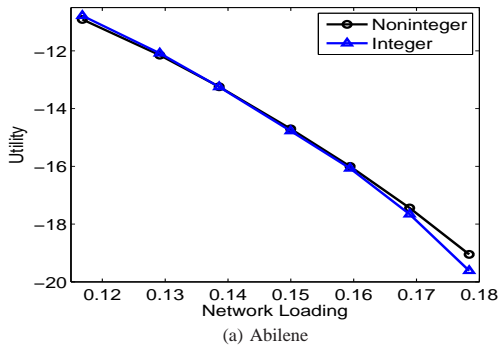}\\
\scriptsize{(a) Abilene}
\end{minipage}
\qquad
\begin{minipage}[t]{8cm}
\centering
\includegraphics[width=7cm,height=4.5cm]{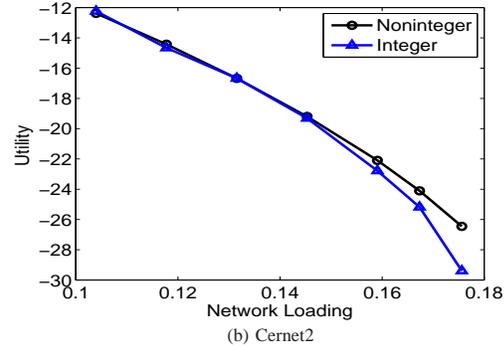}\\
\scriptsize{(b) Cernet2}
\end{minipage}
\renewcommand{\figurename}{Fig.}
\caption{Impact of integer weights on performance}
\label{fig:IntergerLinkWeight}
\end{figure*}

Fig.\ref{fig:IntergerLinkWeight} shows the impact of integer weights on utility for Abilene network and Cernet2 network.
Observe that the integer weights has little impact on utility for the low network loading.  At higher network loadings,
errors due to integer tolerances comes into play so that the utility starts to deviate significantly. This is because, the first link weights will be increasing with the link load while the tolerance for the Dijkstra's algorithm is specified. In order to avoid problems due to such errors, we use the different tolerance for the Dijkstra's algorithm with the different network loading.

\section{Related work}\label{sec:related}

Among the papers focused on TE, MLU \cite{Wang} and piecewise-linear
approximation of the $M/M/1$ delay formula \cite{Fortz00} are two
frequently-used cost functions. Minimizing MLU ensures that the traffic
is moved away from congested hot spots to
less utilized parts of the network. The latter formula proposed by Fortz et al.\cite{Fortz00}
is based on discussions with the technicians in AT\&T Lab.
A description of the general infrastructure behind this
kind of traffic engineering is given in \cite{Fortz02b}. Srivastava
et al. \cite{Srivastava} constructed a composite cost function which
was a positive linear combination of the used capacity and MLU, and
then proposed a heuristic hybrid method combining the
sub-gradient projected method and a genetic algorithm to determine
the link weight system. Di Yuan \cite{BiCriteria} proposed an approach for robust OSPF routing using an artificial objective function embedded into a local search algorithm.

In the research group of congestion control, researchers are mainly
concerned with fairness and efficiency. Network utility maximization (NUM)
\cite{KMT98}, especially the proportionally fair, is a trade-off objective
for this aim. In addition, to design the
end-to-end algorithms for joint routing and rate control, many following
researches replace capacity constraints with barrier functions that
specify the congestion cost at the link (e.g., \cite{KMT98,KellyVoice}
and \cite{HanSrikant}). Generally, a function
$\Phi_{ij}(f_{ij})=\int^{f_{ij}}_0p_{ij}(u)\indif u$ is defined,
which can be regarded as a penalty function that describes the rate
at which the cost is incurred at resource $(i,j)$ with capacity $c(i,j)$ when the load
through it is $f_{ij}$. He et al. \cite{Jia07} choose
$\Phi(f_{ij},c_{ij})=e^{\frac{f_{ij}}{c_{ij}}}$ to model $M/M/1$
queuing delay, which is related to the price function
$p_{ij}(f_{ij})=\frac{1}{c_{ij}}e^{\frac{f_{ij}}{c_{ij}}}$. Xu et
al. in \cite{LBMP} defined
$\Phi_{ij}(f_{ij})=-q_{ij}\ln(c_{ij}-f_{ij}),$ which is related to
the price function $p_{ij}(f_{ij})=\frac{q_{ij}}{c_{ij}-f_{ij}}$.

Based on the results derived from linear programming, Wang et al.\cite{Wang}
proved that any arbitrary set of routes can be converted to
shortest-paths with respect to some set of positive link weights.
This implies that the shortest path limitation is in itself
not a major hurdle. Fortz et al.\cite{Fortz04} showed that
optimizing the link weights for OSPF with evenly split over ECMP to the offered
traffic is an NP-hard problem and proposed a local search heuristic.
Sridharan et al.\cite{Sridharan} used a centralized
greedy computation to select the subset of next-hops for each prefix
to attain load balance much better than even splitting among the
shortest paths. But these solutions fail to enable routers to
independently compute the flow-splitting ratios only using link weights.
PEFT, recently proposed by Xu et al. \cite{PEFT}, is a promising link-state routing
protocol splitting traffic over multiple paths with an exponential
penalty on longer paths. In order to prevent loops and promote computational efficiency, PEFT used
\emph{Downward PEFT} for traffic splitting, which does not provably
achieve optimal TE \cite{PEFT}. PEFT shed a new light for further studies
on developing an OSPF-based protocol, which enables routers to independently make
traffic split decisions while maintaining the shortest paths.

\section{Conclusions}\label{sec:conclusion}
In this paper, we explore the problem of achieving the optimal traffic engineering in intra-domain IP networks. Firstly, we propose a new generic objective function, where various interests of providers can be extracted with different parameter settings. And then, we model the optimal TE as the utility maximization of multi-commodity flows and theoretically show that any given set of optimal routes corresponding to a particular objective function can be converted to shortest paths with respect to a set of positive link weights, which can be directly configured on OSPF-based protocols. On these bases, we develop a new OSPF-based routing protocol, SPEF, to realize a flexible way that splits traffic over shortest paths in a distributed fashion. The inspiring fact lies that comparing to OSPF, SPEF only needs one more weight for each link and provably achieves optimal TE. Numerical experiments have been done to compare SPEF with the current version of OSPF, showing the effectiveness of SPEF in terms of link utilization and network load distribution.

A direction for further studies is that we should analyze the computational complexity in network environment with OSPF as well as other existing approaches including PEFT.

%


\end{document}